\documentclass[11pt, A4]{article}
\usepackage{bbold}

\usepackage{amsfonts}

\usepackage{cancel}
\usepackage{cite}

\usepackage{graphicx}
\usepackage{epstopdf}
\usepackage{amsmath}
\usepackage{amssymb}
\usepackage{amsthm}
\usepackage[usenames]{color}
\usepackage{array}

\usepackage[
      colorlinks=true,
      linkcolor=blue,
      urlcolor=blue,
      filecolor=blue,
      citecolor=red,
      pdfstartview=FitV,
      pdftitle={},
      pdfauthor={},
      pdfsubject={},
      pdfkeywords={},
      pdfpagemode=None,
      bookmarksopen=true
]{hyperref}

\usepackage{epsfig}
\usepackage{mathtools}

\usepackage{hyperref}
\usepackage{comment}

\def\cC{\mathcal{C}}

\def\cM{\mathcal{M}}

\def\d{\partial}

\def\half{\frac{1}{2}}
\def\quart{\frac{1}{4}}

\def\a{\alpha}

\def\e{\epsilon}

\def\G{\Gamma}

\def\r{\rho}
\def\o{\omega}

\def\t{\tau}

\def\z{\zeta}
\def\be{\begin{equation}}
\def\ee{\end{equation}}
\def\bea{\begin{eqnarray}}
\def\eea{\end{eqnarray}}

\newcommand{\rom}[1]{\mathrm{#1}}

\def\cC{\mathcal{C}}

\def\cM{\mathcal{M}}

\def\eps{\epsilon}

\def\no{\nonumber}

\def\nn{\nonumber}

\begin{document}
  
  \begin{centering}


{\LARGE \textsc{ Black Hole Hair Removal  For \\ \vskip 0.3cm N = 4 CHL Models 
}} \\

 \vspace{0.8cm}

{Subhroneel Chakrabarti$^{a,e}$, Suresh Govindarajan$^b$, \\
P Shanmugapriya$^c$, Yogesh K.~Srivastava$^{d,e}$,  \\
and Amitabh Virmani$^c$}
\vspace{0.8cm}

\begin{minipage}{.9\textwidth}\small  \begin{center}
${}^{a}${The Institute of Mathematical Sciences,  CIT Campus, \\
Taramani, 
Chennai, 
Tamil Nadu, India 600113}\\
  \vspace{0.5cm}
$^b$Department of Physics,  Indian Institute of Technology Madras, \\
Chennai,  India 600036 \\
  \vspace{0.5cm}
$^c$Chennai Mathematical Institute, H1 SIPCOT IT Park, \\ Kelambakkam, Tamil Nadu, India 603103\\
  \vspace{0.5cm}
${}^{d}${National Institute of Science Education and Research (NISER), \\ Bhubaneswar, P.O. Jatni, Khurda, Odisha, India 752050}\\
  \vspace{0.5cm}
$^e$Homi Bhabha National Institute, Training School Complex, \\ Anushakti Nagar, Mumbai, India 400085 \\
  \vspace{0.5cm}
{\tt subhroneelc@imsc.res.in, suresh@physics.iitm.ac.in, shanmugapriya@cmi.ac.in,\\ yogeshs@niser.ac.in, avirmani@cmi.ac.in}
\\ $ \, $ \\

\end{center}
\end{minipage}

\end{centering}

\begin{abstract}
Although BMPV black holes in flat space and in Taub-NUT space have identical near-horizon geometries, they have different  indices from the microscopic analysis.  For K3 compactification of type IIB theory, Sen et al in a series of papers identified that the key to resolving this puzzle is the black hole hair modes: smooth, normalisable,  bosonic and fermionic degrees of freedom living outside the horizon.  In this paper, we extend their study to N = 4 CHL orbifold models. For these models,  the puzzle is more challenging due to the presence of the twisted sectors. We  identify  hair modes in the untwisted as well as twisted sectors. We show that  after removing the contributions of the hair modes from the microscopic partition functions,  the 4d and 5d horizon partition functions  agree. Special care is taken to present details on the smoothness analysis of hair modes for rotating black holes, thereby filling an essential gap in the literature. 
 \end{abstract}

\newpage

\tableofcontents

\newpage

\setcounter{equation}{0}
\numberwithin{equation}{subsection}

\section{Introduction}

For a class of supersymmetric black holes in  
five- and 
four-dimensional theories of gravity, string theory explains the entropy in terms of underlying microscopic degrees of freedom. Earlier studies dealt with black holes carrying large charges~\cite{Strominger:1996sh}, and found that in the large charge limit string theory gives a perfect match with the Bekenstein-Hawking entropy of the corresponding black hole in the two-derivative theory. This matching is widely regarded as one of the biggest successes of string theory.

 Since then, this matching has been improved from both the gravity and the microscopic sides.  On the microscopic side, for a class of black holes, precision counting formulas are known~\cite{hep-th/9607026, Maldacena:1999bp, hep-th/0503217, hep-th/0505094, 0708.1270}. These counting formulas, also often known as the black hole partition functions,  give an integer for the indices corresponding to the  BPS microstates underlying the given supersymmetric black hole. On the gravity side, Sen's quantum entropy function formalism posits that a  path integral computation in string theory on $\mathrm{AdS}_2 \times \mathrm{K}$ near-horizon geometry should give the black hole indices in the full quantum theory~\cite{Sen:2008vm, Mandal:2010cj, Sen:2014aja},  where $\mathrm{K}$ is a compact manifold.  Since this computation only refers to the  near-horizon geometry, the answer is expected not to be sensitive to the nature of the solution far away from the horizon. That is, if two black holes have identical near-horizon geometries, they must have identical microscopic indices. 

There is, however, a well known and well-studied counterexample to this: the BMPV black hole in flat space~\cite{Breckenridge:1996is} versus the BMPV black hole in Taub-NUT space~\cite{Gauntlett:2002nw, hep-th/0503217}.   These two types of black holes have identical near-horizon geometries but different  microscopic indices. Banerjee, Mandal, Jatkar, Sen, and Srivastava (one of the authors of this paper) in \cite{Banerjee:2009uk, Jatkar:2009yd} identified that the key to the resolution of this puzzle is the black hole hair modes: smooth, normalisable,  bosonic and fermionic degrees of freedom living outside the horizon.  For the case of $\text{K3}$ compactification of type IIB theory, Jatkar, Sen, and Srivastava~\cite{Jatkar:2009yd} constructed hair modes as non-linear solutions to the supergravity equations and showed that once the contributions of the hair modes are properly removed, the 4d and 5d partition functions match.

The purpose of this paper is to extend the non-linear hair mode analysis of~\cite{Jatkar:2009yd}  in two ways:
 \begin{itemize}
 \item In \cite{Jatkar:2009yd}, smoothness analysis for hair modes was mostly presented for non-rotating black holes, even though rotation plays an important role in their arguments.  This paper presents details on the smoothness analysis of hair modes for rotating black holes, thereby filling this gap in the literature.  The smoothness analysis is essential. For example, in reference  \cite{Banerjee:2009uk}, certain modes corresponding to the transverse oscillations of black holes were counted as hair modes but were later removed from the counting in \cite{Jatkar:2009yd}, as these modes turn out to be singular at the horizon of the black hole.  To confirm that rotation does not alter any of the conclusions of~\cite{Jatkar:2009yd}, it is important to fill this gap. 
  
 \item  We extend the matching of hair removed partition functions to more general  $\text{K3}$ compactifications, the so-called Chaudhuri-Hockney-Lykken (CHL) orbifold models \cite{Chaudhuri:1995fk, Chaudhuri:1995bf, Schwarz:1995bj, Chaudhuri:1995dj}. These models are widely studied in the context of precision counting of black hole microstates~\cite{hep-th/0510147, hep-th/0602254, hep-th/0605210, hep-th/0609109,  Sen:2007vb, Dabholkar:2007vk,  Cheng:2007ch, Dabholkar:2008zy,  0907.1410, 1109.3706, Chowdhury:2019mnb, Fischbach:2020bji, Cardoso:2020swg} and are reviewed in \cite{0708.1270}. For these models,  the puzzle of the mismatch of the 4d and 5d partition functions
is more challenging due to the presence of the twisted sectors. We  identify  hair modes in the untwisted as well as twisted sectors and show that  after removing the contributions of the hair modes from the microscopic partition functions, the 4d and 5d horizon partition functions perfectly match.
 \end{itemize}

The rest of the paper is organised as follows. 

All of our analysis in sections \ref{sec:sugra} through \ref{sec:deformations_TN} is in six-dimensional (2,0) supergravity coupled to $n_t$ number of tensor multiplets. 

In section
\ref{sec:sugra}, we present relevant details on (2,0) supergravity coupled to $n_t$ number of tensor multiplets. Most studies on the CHL models are done in four dimensions. To facilitate the transition between our  six-dimensional notation and a four-dimensional notation,  we recall that  upon dimensional reduction, each self-dual or anti-self-dual tensor in six-dimensions gives a vector in four dimensions. From the pure 6d (2,0) graviton multiplet, we get 7  vectors in 4d: 5 from the self-dual tensor fields and 2 as graviphotons. Hence the number of vectors in 4d is, 
\be
n_V^{\rom{4d}} = n_t + 7.
\ee
Often a notation $n_V^{\rom{4d}} = 2k + 8$ is used. Thus,
\be
n_t = 2 k + 1 \implies k+2 =  \frac{1}{2} (n_t + 3).
\ee
The number $n_t + 3$ plays a key role in our later considerations.

In
section
\ref{sec:black_hole_BMPV}, BMPV black hole in flat space is discussed. The near-horizon geometry, a set of vielbeins, and Killing spinors are presented. A set of coordinates are
introduced in which the metric and the three-form field strength are analytic near the horizon.

In section \ref{sec:deformations}, hair modes  on BMPV black hole are studied. Specifically, 
in section \ref{sec:bosonic_deformations} bosonic deformations of the BMPV black hole generated by the Garfinkle-Vachaspati transform are studied.   It is found that all these bosonic deformations are non-smooth. In section  \ref{sec:fermionic_deformations} fermionic deformations of the BMPV black hole are studied. 

In section \ref{sec:black_hole_BMPV_TN}, BMPV black hole in Taub-NUT space is discussed. As in the earlier section on BMPV black hole in flat space, the near-horizon geometry, a set of vielbeins, and Killing spinors are presented. A set of coordinates are introduced in which the metric and  the three-form field strength  are analytic near the horizon. 

Hair modes on BMPV black hole in Taub-NUT are studied in section \ref{sec:deformations_TN}. In section \ref{sec:bosonic_deformations_TN}, bosonic deformations of the BMPV black hole in Taub-NUT generated by the Garfinkle-Vachaspati transform are studied. In section \ref{sec:bosonic_deformations_TN_form_fields}, a class of deformations corresponding to anti-self-dual form fields are studied. In section  \ref{sec:fermionic_deformations_TN}, fermionic deformations are studied. A key observation of this section is that the number of hair mode deformations corresponding to the anti-self-dual form fields is equal to the number of tensor multiplets $n_t$ in the 6d theory. Since different CHL models have a different number of tensor multiplets, the number of such (untwisted sector) hair modes change from theory to theory.

In section \ref{sec:microscopics},
we turn to the discussion of hair removed 4d and 5d partition functions. In section \ref{4d_5d_counting_formulas}, we first review the microscopic considerations relevant for our discussion and highlight that the 4d and 5d partition functions (and  microscopic degeneracies) are different. Due to the presence of twisted sectors in type IIB CHL models, the difference in  the 4d and 5d partition functions is quite non-trivial.

In section \ref{hair_partition_functions}, we identify twisted sector hair modes in ten-dimensional supergravity description, and compute the hair removed 4d and 5d partition functions.  The twisted sector hair modes can be schematically  understood as follows. We recall that the CHL models are obtained as  $\mathbb{Z}_N$ orbifold of type IIB theory on $\text{K3} \times \text{S}^1 \times \widetilde{\text{S}}^1$.  The orbifold group is generated by $\widetilde{g}$ such that $\widetilde{g}^N=\mathbb{1}$. The orbifold action also involves a shift along the S$^1$.  To obtain the six-dimensional supergravity description, only the $\widetilde{g}-$invariant fields are kept. Hence in six-dimensions, we only see the $\widetilde{g}-$invariant (or the untwisted sector)  hair modes.  In ten-dimensions,  a more general situation is possible. Let $C^\rom{10d}_4$ denote the ten-dimensional RR four-form field. Let $\omega$  denote a two-form in the cohomology of K3 that is not $\widetilde{g}-$invariant. Then, it can be multiplied with a two-form $c_{2}^\rom{6d}$ (not visible in 6d supergravity)
 \be 
 C^\rom{10d}_4 \propto  c^\rom{6d}_{2} \wedge \omega^\rom{K3} ,
\ee
such that $c^\rom{6d}_{2}$ picks up the \emph{opposite}  phase under the orbifold action compared to $\omega^\rom{K3}$. The combined effect ensures that the ten-dimensional $ C^\rom{10d}_4$ is $\widetilde{g}-$invariant. These modes give rise to additional hair modes.  In section \ref{matching},
we show that the hair removed 4d and 5d partition functions perfectly match. 

We conclude with a summary and a brief discussion of open problems in section \ref{sec:conclusions}.

  The main body of the paper is mostly a technical analysis of either the hair modes or the microscopic partition functions. Some additional technical details are relegated to  two appendices.  
The authors of \cite{Jatkar:2009yd} introduced the concept of weight that proves very convenient, both in the analysis of the background solutions and  in the analysis of  bosonic and fermionic deformations. In appendix \ref{app:weights} we review the concept of weight. In the smoothness analysis of the hair modes,   we find that often a specific set of Lorentz transformations needs to be performed on local Lorentz frames in six-dimensions. The transformation of the gravitino field under those Lorentz transformations is discussed in appendix \ref{app:Lorentz}.

Readers only interested in the microscopic analysis may choose to skip directly to section \ref{sec:microscopics}.  
 In order to help such readers, we now briefly summarise the \emph{untwisted} sector hair modes for both types of black holes found in sections \ref{sec:deformations} and \ref{sec:deformations_TN}, respectively. The twisted sector hair modes are discussed in section \ref{sec:microscopics}.

For the  BMPV solution, the six directions are the four transverse spatial directions together with an S$^1$ and the time direction. The S$^1$  is along which the D1 and D5 branes wrap in the brane description of the BMPV solution. Since the BMPV solution is independent of the S$^1$ coordinate (labelled $x^5$), it is useful to regard it as a black string extended along the S$^1$ direction. In such a description, a left-moving mode represents a set of deformations labelled by an \emph{arbitrary} function of the light cone coordinate $v: = x^5 + t$. Functions of coordinate $v$ describe the propagation of plane waves along the negative S$^1$ direction.  Geometric quantisation of these modes is expected to generate degeneracies associated with hair modes. 

For the four-dimensional black hole, i.e., a BMPV black hole in  Taub-NUT space, the same discussion applies except the four transverse spatial directions are now the four  directions of the Taub-NUT space. The Taub-NUT circle is denoted as $\widetilde{\mathrm{S}}^1$. Dimensional reduction over $\mathrm{S}^1 \times \widetilde{\mathrm{S}}^1$ gives the four-dimensional description.

For the five-dimensional BMPV black hole, the hair modes consist of (apart from zero modes):
\begin{itemize}
\item Four left-moving fermionic modes describing the propagation of goldstino modes associated with four of the twelve broken supersymmetries.
\end{itemize}
In reference  \cite{Banerjee:2009uk} four left-moving bosonic modes corresponding to the transverse oscillations of the BMPV black string were also counted but were later removed from the counting in \cite{Jatkar:2009yd}, as these modes turn out to be singular at the horizon of the black hole. We show that the same conclusion holds for the rotating black holes.

For the BMPV black hole in Taub-NUT space, the hair modes consist of (apart from zero modes) in 6d supergravity analysis:
\begin{itemize}
\item Four left-moving fermionic modes describing propagation of goldstino modes associated with four of the twelve broken supersymmetries, as in the BMPV case.
\item $n_t$ left-moving bosonic modes arising from certain deformations of the anti-self-dual two form field of each tensor multiplet.
\item Three left-moving bosonic modes describing deformations in the three non-compact directions of the Taub-NUT space. 
\end{itemize}
In reference  \cite{Banerjee:2009uk} four left-moving bosonic modes corresponding to the transverse oscillations of the BMPV black string relative to the Taub-NUT were also counted but were later removed from the counting in \cite{Jatkar:2009yd}.   At the black hole horizon, these modes are expected to be exactly the same as those for the BMPV black hole, and hence  singular.  Although an explicit analysis of these modes is likely to be not difficult, it is still missing in the published literature.\footnote{We thank Samir Mathur and Dileep Jatkar for discussions on this point.}

As an important summary point, we note that the 4d black hole has $n_t + 3$ additional left-moving bosonic hair modes compared to the 5d black hole. The number $n_t + 3$ plays a key role in section \ref{sec:microscopics}. 

\paragraph{Note Added:} A few days after this paper appeared on the arXiv, a paper by Chattopadhyaya and David \cite{Chattopadhyaya:2020yzl} also appeared that overlaps with our section~\ref{sec:microscopics}. Specifically, our counting of hair modes, boundary conditions analysis of hair modes on the $\mathrm{S}^1$ and hair removed partition functions overlap with the results of \cite{Chattopadhyaya:2020yzl}.

\numberwithin{equation}{subsection}

\section{Supergravity set-up}

\label{sec:sugra}

We now present relevant details on (2,0) supergravity coupled to $n_t$ number of tensor multiplets. Such a theory is obtained from appropriate truncation of type IIB theory on K3/$\mathbb{Z}_N$ or $\mathrm{T}^4$/$\mathbb{Z}_N$ with a total of 16 supersymmetries. The case of $n_t =21$ corresponds to K3 compactification considered in \cite{Jatkar:2009yd}. We present the discussion in two steps: (i) a truncation of IIB theory compactified on $\text{T}^4$ to pure (2,0) 6d supergravity, (ii) coupling this $(2,0)$ theory to  $n_t$ tensor multiplets. This split is artificial, but we found it easiest to think in these terms.

Toroidal reduction of type IIB supergravity to six dimensions leads to the unique six-dimensional $(2, 2)$ supergravity.  The spectrum of $(2, 2)$ six-dimensional supergravity consists of a graviton, 8 gravitinos, 5 two-forms, 16 gauge fields, 40 fermions,  and 25 scalars. See, for example, table 5 of  \cite{deWit:2002vz}. This theory was first constructed by Tanii \cite{Tanii:1984zk}. It can be consistently  truncated to pure $(2, 0)$ supergravity by setting the 4 right chiral gravitinos, anti-self-dual parts of the 5 two-forms fields, 16 gauge fields, 40 fermions, and 25 scalars to zero \cite{Tanii:1984zk}. In other words, the resulting $(2,0)$ theory consists of a graviton, 4
gravitinos, and 5 self-dual tensor fields. 

A tensor multiplets in six-dimensions contains an anti-self-dual
tensor field, 4 fermions and 5 scalars. If we consider pure $(2,0)$ theory coupled to $n_t$ tensor multiplets, then we have the field content: 
a graviton, 4 left chiral gravitinos, 5 self-dual two-forms, $n_t$ anti-self-dual two-forms, $4 n_t$ fermions, and $5n_t$ scalars. For $n_t = 5$ this is a consistent truncation of $(2, 2)$ supergravity. In this truncation, compared to the $(2,2)$ theory,  we have left out the 4 right chiral gravitinos, 16 gauge fields, and 20 fermions.  This truncation is often used in the studies of the D1-D5 system \cite{Bossard:2019ajg}.

For the $(2,0)$ theory coupled to $n_t$ tensor multiplets, we follow the conventions of \cite{Deger:1998nm} together with the simplifications introduced in \cite{Jatkar:2009yd}. In the following, we summarise the relevant bosonic/fermionic equations of motion and the Killing spinor equations, restricting ourselves to the details we need  later. A complete description can be found in \cite{Deger:1998nm}.

Let us denote the
self-dual and the anti-self-dual field strengths by $\bar H^k_{MNP}$
($1\le k\le 5$) and $H^s_{MNP}$ ($6\le s\le n_t + 5)$ respectively,
satisfying
\bea 
\label{eself} \bar H^{kMNP} &=& + \frac{1}{3!} \, |\det
g|^{-1/2} \epsilon^{MNPQRS}\, \bar H^k_{QRS}, \\
\label{eanti_self} H^{sMNP} &=& -\frac{1}{3!} \, |\det g|^{-1/2} \epsilon^{MNPQRS}\, H^s_{QRS} \, ,
\eea
where $\epsilon^{MNPQRS}$ is the totally anti-symmetric symbol. 
The sign convention for the anti-symmetric symbol  will be made explicit in the next section.  Throughout this paper, we
shall set all the scalar fields to fixed (attractor)
values. As a result, all the derivatives of the scalar fields are zero. This significantly simplifies the presentation of the equations of motion. For the bosonic equations, we then have
\be
\label{eeom1}
 R_{MN} = \bar H^k_{MPQ} \, \bar H_N^{kPQ} + H^s_{MPQ} \, H_N^{sPQ},
\ee
together with
\be
\label{eeom2}
\bar H^k_{MNP} H^{sMNP}  = 0. 
\ee
Following \cite{Jatkar:2009yd} we choose the convention where $\bar H^1$ and $H^6$ denote
the self-dual and anti-self-dual components of the six-dimensional  $F^{(3)}_{MNP}$ Ramond-Ramond (RR) field coming from  IIB theory. The six-dimensional version of the 3-form RR field is obtained by simply restricting the indices of the ten-dimensional RR field to six-dimensions. More precisely, 
\be \label{ef3}
F^{(3)}_{MNP} = 2\, e^{-\Phi}\,
\left(\bar H^1_{MNP} + H^6_{MNP}\right)\, ,
\ee
where the dilaton $\Phi$ takes a constant (attractor) value in the backgrounds we consider.

The fermion fields in  six-dimensional theory consist of a set of four left-chiral
gravitinos $\Psi_M^\alpha$ ($0\le M\le 5$, $1\le\alpha\le 4$, spinor indices are suppressed) and
a set of $4 n_t$ right-chiral spin-1/2 fermions $\chi^{\alpha
r}$ ($1\le r\le n_t$, for $n_t$ tensor multiplets). Let $\Gamma^M$'s  $(0\le M\le 5)$
denote $8\times 8$ gamma matrices for six spacetime dimensions written
in the \emph{coordinate} basis. Let $A, B, \ldots$ denote the six-dimensional tangent
space indices, so that
\be
e^{A}_M \Gamma^M
\ee
are the standard six-dimensional Clifford algebra matrices. In order to avoid any notational confusion, we put wide-tildes on the tangent space gamma matrices and define,
\be
\widetilde \Gamma^A = e^{A}_M \Gamma^M. 
\ee
The Clifford algebra is then, 
\be 
\{\widetilde\Gamma^A,  \widetilde\Gamma^B\}=2 \eta^{AB}, 
\ee
for $0 \le A, B \le 5$ tangent space indices. 
With this notation, the chirality conditions for the fermionic fields are 
\be \label{gravchiral}
\left(\frac{1}{6!} \, |\det g|^{-1/2}\,
\epsilon^{MNPQRS} \Gamma_{MNPQRS} +1\right)\Psi_M^\alpha
=0 \quad \implies \quad (\widetilde \Gamma_{012345} +1) \, \Psi_M^\alpha
=0\, ,
\ee
\be \label{chichiral}
\left(\frac{1}{6!} \, |\det g|^{-1/2}\,
\epsilon^{MNPQRS} \Gamma_{MNPQRS} -1\right)\chi^{\alpha r}
=0 \quad \implies \quad (\widetilde \Gamma_{012345} -1) \, \chi^{\alpha r}
=0.
\ee

The R-symmetry group of  $(2,0)$ theory is SO(5) $\simeq$ USp(4). Following \cite{Deger:1998nm}, we exclusively work with SO(5) notation.\footnote{Reference \cite{Bossard:2019ajg} works with the USp(4) notation. }  In this notation,  $\alpha, \beta, \ldots$ are  spinor indices of  SO(5) and $i,j,k, \ldots$  are  vector indices of SO(5) (e.g., the index $k$ used above for the self-dual tensors). In order to work with SO(5) spinor indices we also need to introduce $4 \times 4$ SO(5) gamma matrices. We denote these matrices with wide-hats $(\widehat \Gamma^i)_{\alpha \beta}$. They satisfy Euclidean Clifford algebra in five-dimensions, 
\be
\{\widehat\Gamma^k, \widehat\Gamma^l\} =2\delta_{kl}.
\ee

In the fermionic sector, our considerations are restricted to linear equations of motion in the spinor fields (at least to begin with). 
For the backgrounds we will work with, not only the scalars are set to constant values but also the spin-1/2 fermions $\chi^{\alpha r}$ are all set to zero. With these conditions, the fermion sector field equations of motion simplify to 
\bea
& & \Gamma^{MNP} D_N\Psi^\alpha_P   - \bar H^{kMNP} \Gamma_N\, \widehat \Gamma^k_{\alpha\beta}   \Psi^\beta_P= 0,\\
&& H^{sMNP} \, \Gamma_{MN} \Psi^\alpha_P=0,
\eea
where the $D_N$ is the standard full covariant derivative with the unique torsion free spin connection. In complete detail, 
\be 
D_M\Psi^\alpha_P = \partial_M \Psi^\alpha_P - \Gamma^N_{MP}
\Psi^\alpha_N + \quart \,
\omega_{MAB} \, \widetilde \Gamma^{AB}\,  \Psi^\alpha_P\, ,
\ee
where
\be  
\Gamma^M_{NP} \equiv \half \, G^{MR}\,
(\partial_N G_{PR} + \partial_P G_{NR} -\partial_R G_{NP}),
\ee
and
\be
\omega_M^{AB} \equiv 2 e^{N[A}\partial_{[M} e_{N]}^{B]} - e^{N[A}e^{B]P}e_{MC}\partial_N e_P^C,
\ee
where we  remind the reader that $M, N, P, \ldots$ are the spacetime indices and $A, B, C \ldots$ are the tangent space indices. For the first order gravity manipulations we follow the standard conventions, e.g., chapter 7 of \cite{Freedman:2012zz}.

We now discuss the Killing spinor equations.  For the black hole backgrounds \emph{without} hair the anti-self-dual fields are all zero. With this simplification the Killing spinor equations reduce to only one equation, 
\be
D_M \epsilon - \frac{1}{4} \bar{H}^i_{MNP} \Gamma^{NP} \widehat \Gamma^i \epsilon = 0.
\label{Killing_Spinor}
\ee
In writing this equation all spinor indices are suppressed.  $\epsilon$ is the supersymmetry transformation parameter. Since the theory is a chiral theory, the  supersymmetry parameter satisfies, 
\be
(\widetilde \Gamma_{012345} +1) \epsilon = 0,
\ee
i.e., $\epsilon$ is a six-dimensional left chiral Weyl spinor. In six-dimensions the symplectic Majorana condition is consistent with chirality, so in addition $\epsilon$ satisfies,
\begin{align}
  \bar \epsilon &= \epsilon^T \, C \, \Omega, &  \Omega^T &= - \Omega,
\end{align}
where $C$ is the symmetric charge conjugation matrix for the six-dimensional Clifford algebra for the Lorentz group SO(5,1) and $\Omega$ is the anti-symmetric charge conjugation matrix for the Euclidean five-dimensional Clifford algebra for the R-symmetry group SO(5). $\Omega$ being antisymmetric is the only consistent choice for Euclidean five-dimensional space, see e.g., Table 1 of \cite{VanProeyen:1999ni}. Moreover, 
\be
(C \widetilde \Gamma^A)^T = - C \widetilde \Gamma^A, \qquad(\Omega \widehat \Gamma^i)^T = - \Omega \widehat \Gamma^i.
\ee

To discuss supersymmetry of black holes \emph{with} hair, we will also need the Killing spinor equations  when the gravitino fields and the anti-self-dual fields are not set to zero.  The equations are obtained by setting the supersymmetry variation of all fields to zero.  For  the 
vielbein, the gravitinos, the self-dual 2-form fields and the spin-1/2 fields  respectively\footnote{For the backgrounds of interest, the supersymmetry variation of the anti-self-dual fields and scalars  do not give non-trivial equations.}, these equations are~\cite{Deger:1998nm}, 
\bea
& & \bar \epsilon \, \widetilde \Gamma^A \Psi_M = 0, \label{del_metric} \\
& & D_M \epsilon - \frac{1}{4} \bar{H}^i_{MNP} \Gamma^{NP} \widehat \Gamma^i \epsilon = 0, \label{del_grav} \\
& & \bar \epsilon \, \Gamma_{[M} \widehat \Gamma^i \Psi_{N]} = 0, \label{del_FF} \\
& & \Gamma^{MNP} H^{s}_{MNP} \epsilon = 0. \label{del_spin1/2}
\eea

\numberwithin{equation}{subsection}
\section{BMPV black hole in flat space}

\label{sec:black_hole_BMPV}

In this section, we start by reviewing the BMPV black hole \cite{Breckenridge:1996is} in flat space. In section \ref{sec:metric} various coordinates we need to describe the black hole, metric, three-form field strength, and near-horizon geometry are presented. In section \ref{sec:killing_spinors} Killing spinors for these black holes are constructed. In section \ref{sec:non_singular_coordindate}   coordinates  in which the black hole metric is smooth at the future horizon are constructed.

\subsection{Coordinates, metric, and form field}

\label{sec:metric}

In a standard set of coordinates four-dimensional Euclidean flat space takes the form
\be
ds^2 = d \tilde r^2 + \tilde r^2 (d\tilde \theta^2 + \cos^2\tilde \theta d \tilde \phi^2 +  \sin^2\tilde \theta d \tilde \psi^2).
\ee
These coordinates are related to cartesian coordinates as, 
\begin{align}
w^1 &= \tilde r \cos \tilde \theta \cos \tilde \phi, &
w^2 &= \tilde r \cos \tilde \theta \sin \tilde \phi,\\
w^3 &= \tilde r \sin \tilde \theta \cos \tilde \psi ,&
w^4 &= \tilde r \sin \tilde \theta \sin \tilde \psi.
\end{align}
To cover the full range of $w^i$ we need to restrict ourselves to,
\be
\tilde \theta \in \left (0, \frac{\pi}{2}\right), \qquad \tilde \phi \in (0, 2 \pi), \qquad \tilde \psi \in (0, 2\pi), \label{ranges}
\ee
with
\be
(\tilde \phi, \tilde \psi ) \equiv (\tilde \phi + 2 \pi, \tilde \psi ) \equiv (\tilde \phi, \tilde \psi  + 2 \pi).\label{identifications}
\ee
We can extend the range of $\tilde\theta$ to $(0, \pi)$ by introducing the identification,
\be
(\tilde\theta, \tilde\phi, \tilde\psi) \equiv (\pi - \tilde \theta, \tilde\phi + \pi , \tilde\psi). \label{identifications2}
\ee

The coordinates we will use for the most part are the Gibbons-Hawking coordinates $(r,\theta, \phi, x^4)$ defined via,
\begin{align}
\tilde r &= 2 \sqrt{r}, &
\tilde \theta &= \frac{\theta}{2},\\
 \tilde \phi &= \frac{1}{2} (x^4 + \phi), &
 \tilde \psi &= \frac{1}{2} (x^4 - \phi). 
 \end{align}
In these coordinates flat space metric becomes,
\be
ds^2_\rom{flat} = \frac{1}{r}dr^2 + r (d\theta^2 + \sin^2 \theta d\phi^2 + (dx^4+\cos \theta d\phi)^2),
\ee
and the identifications \eqref{identifications}-\eqref{identifications2} become, 
\be
(\theta, \phi, x^4) \equiv (2\pi - \theta, \phi + \pi, x^4 + \pi) \equiv (\theta, \phi+2\pi, x^4+2\pi)\equiv (\theta, \phi, x^4+ 4 \pi).
\ee

Next we introduce a one-form
\be
\chi_i dw^i = - 2 \zeta, 
\ee
where
\be
 \zeta = - \frac{\widetilde J}{8 r} (dx^4 + \cos \theta d\phi).
\ee
In the other coordinates introduced above $\zeta$ takes the form, 
\bea
\zeta
 = - \frac{\widetilde J}{\tilde r^2} (\cos^2 \tilde \theta d\tilde \phi + \sin^2 \tilde \theta d\tilde \psi)
= -\frac{\widetilde J}{\tilde r^4} \left(w^1 dw^2 - w^2 dw^1 + w^3 dw^4 - w^4 dw^3\right).
\eea

The BMPV black hole metric takes the form,
\bea
ds^2 &=& G_{MN} dx^M dx^N \nonumber \\
&=& \psi^{-1}(r) \left[ du dv + (\psi(r)-1)dv^2  -2 \zeta dv \right] + \psi(r) ds^2_\rom{flat},
\eea
where
\begin{align}
u &= x^5 - t, &v &= x^5 + t.
\end{align}
The  $x^5$ coordinate is periodic with size $2 \pi R_5$. It is the S$^1$ along which the BMPV string is extended. The D1 and D5 branes intersect on this S$^1$ and the momentum is also carried along this direction. 
Often the BMPV black hole is written with three harmonic functions, one each for D1, D5, and P charges. Here we have set them all equal to $\psi(r)$, where
\be
\psi(r) = 1 + \frac{r_0}{r}.
\ee 
The six-dimensional dilaton is set to its constant asymptotic value throughout the spacetime, 
\be
e^{-2\Phi} = \lambda^{-2}. \label{dilaton}
\ee
The only other nontrivial field is the three-form RR field. It takes the form
\be
F^{(3)} = \frac{r_0}{\lambda} \left(\epsilon_3 + \ast_6 \epsilon_3 + \frac{1}{r_0} \psi(r)^{-1} dv \wedge d \zeta\right), \label{F3}
\ee
where  we use the conventions
$
\epsilon^{t54r\theta\phi} = +1
$
and where
\be
\epsilon_3 = \sin \theta \ dx^4 \wedge d\theta  \wedge d\phi.
\ee

The first two terms in \eqref{F3}, $\left(\epsilon_3 + \star_6 \epsilon_3\right)$, are manifestly self-dual as $(\star_6)^2 \epsilon_3 = \epsilon_3$. Performing a small calculation one can see that $\psi(r)^{-1} dv \wedge d \zeta$
is also self dual. Hence the full $F^{(3)}$ is self-dual. Thus, for the BMPV background the anti-self-dual fields are all set to zero. 
Among the self-dual three-form fields, only  $\bar{H}^1_{MNP}$ is non-zero for the BMPV background. It takes the value,
\be
 \frac{1}{3!} \bar H^1_{MNP} dx^M \wedge dx^N \wedge dx^P = \frac{\lambda}{2}F^{(3)} = \frac{1}{2} (1 + \star_6) \epsilon_3 r_0 + \frac{1}{2} \psi(r)^{-1} dv \wedge d \zeta.
\ee
Now, we can readily check that the BMPV background satisfies the bosonic equation of motion, 
\be
R_{MN} = \bar H^1_{MPQ} \bar H^1_{N}{}^{PQ}. 
\ee
The other bosonic equations are trivially satisfied as all the anti-self-dual fields vanish.

To obtain the near-horizon geometry, we work with $t, v, x^4, r, \theta, \phi$ coordinates. We rescale $r$ and $t$ as follows
\be
\label{NHcoord}
r = r_0 \beta \rho, \qquad \qquad t = \tau/\beta,
\ee
and take the limit $\beta \to 0$. We get the near-horizon metric,
\be
\label{NHline}
ds^2 = r_0 \frac{d\rho^2}{\rho^2} + dv^2 + \frac{\widetilde J}{4 r_0} dv(dx^4 + \cos \theta d\phi) - 2 \rho dv d\tau + 4 r_0 d\Omega_3^2, 
\ee
where $d\Omega_3^2$ is the metric on the round three-sphere in Gibbons-Hawking coordinates,
\be
d\Omega_3^2\equiv \quart \left((dx^4+\cos\theta d\phi)^2
+d\theta^2 +\sin^2\theta d\phi^2\right) . 
\label{dOmega3}
\ee 
In this limit the form field becomes,
\be
\label{F3_NH}
F^{(3)} = \frac{r_0}{\lambda} \left[\epsilon_3 + \star_6 \epsilon_3 + \frac{\widetilde J}{8 r_0^2}  dv \wedge \left(\frac{1}{\rho} d\rho \wedge (dx^4 + \cos \theta d\phi) + \sin \theta d\theta \wedge d\phi\right)\right], 
\ee
and the dilaton remains the same, cf.~\eqref{dilaton}.  Note that the $\star_6$ is now with respect to the metric \eqref{NHline}.

For later use we introduce the following vielbeins,
\bea
\label{frame2FS}
e^0 &=& \psi^{-1}(r) (dt+ \zeta), \\
e^1 &=& \left(dx^5+dt- \psi^{-1}(r) (dt+ \zeta) \right), \\
e^2 &=& \psi^{1/2}(r) \, r^{1/2} (dx^4 +\cos\theta d\phi), \label{frame2FS_2}
\\
e^3 &=& \psi^{1/2}(r) \, r^{-1/2} \, dr \, ,  \\
e^4 &=& \psi^{1/2}(r)\, r^{1/2}  \, d\theta \, ,  \\
e^5 &=& \psi^{1/2}(r)\, r^{1/2}  \, \sin\theta\, d\phi\, .
\label{frame2FS5}
\eea
With a bit of calculation one sees that the background fields can be written in terms of the orthonormal frame \eqref{frame2FS}--\eqref{frame2FS5} as
\be
ds^2 = - (e^0)^2 + (e^1)^2 + (e^2)^2 + (e^3)^2
+(e^4)^2 + (e^5)^2,
\label{metric_frame}
\ee
and
\bea
\label{form_field_frame}
F^{(3)} &=& \frac{r_0}{\lambda\, r^2}\,
\Big{[} \psi^{-3/2}(r)\,
r^{1/2}\, (e^2\wedge e^4\wedge e^5
+ e^0\wedge e^1\wedge e^3)   \\ &&
+ \frac{\widetilde{J}}{8 r_0}  \psi^{-2}(r)\,
(-e^0\wedge e^2\wedge e^3 + e^0\wedge e^4\wedge e^5
-e^1\wedge e^2\wedge e^3 + e^1\wedge e^4\wedge e^5)
\Big{]}.\nonumber
\eea
Our convention is $\epsilon^{012345} = +1$. In this form, it can be easily checked that the three-form $F^{(3)}$ is self-dual.

\subsection{Killing spinors}
\label{sec:killing_spinors}
In this section we write explicit expressions for  Killing spinors for the BMPV black string. The Killing spinor equation is \eqref{Killing_Spinor}. Among the self-dual three-form fields, only  $\bar{H}^1_{MNP}$ is non-zero for the BMPV background. As a result the Killing spinor equation   simplifies to,
\be
D_M \epsilon - \frac{1}{4} \bar{H}^1_{MNP} \Gamma^{NP} \widehat{\Gamma}^1 \epsilon = 0.
\ee
We work with $u,v, x^4, r, \theta, \phi$ coordinates.  We demand the projection conditions,
\bea
\Gamma^v \epsilon &=& 0 \label{projection},\\
\widehat{\Gamma}^1 \epsilon &=& \epsilon. \label{projection2}
\eea
Due to projection condition \eqref{projection2}, the Killing spinor equation further simplifies to 
\be
D_M \epsilon - \frac{1}{4} \bar{H}^1_{MNP} \Gamma^{NP}   \epsilon = 0. \label{Killing_Spinor_Simp}
\ee

At this stage, introducing the notion of weight developed in \cite{Jatkar:2009yd} proves convenient.\footnote{This notion of weight is closely  related to the concept of boost weight used in general relativity. It is often used in the algebraic classification of the Weyl tensor based on the existence of preferred null directions. See, e.g.,~\cite{Ortaggio:2012jd}.}  For a more detailed discussion see appendix \ref{app:weights}. The weight is defined for \textit{components} of any tensor in $u$ and $v$ coordinates.   For a given component of a covariant tensor, it is defined as 
\begin{equation}
    wt_{_{cov}} = \# \text{ of } v \text{ indices } - \# \text{ of } u \text{ indices}.
\end{equation}
For a given component of a contravariant tensor, weight is defined as 
\begin{equation}
    wt_{_{cont}} = \# \text{ of } u \text{ indices } - \# \text{ of } v \text{ indices}.
\end{equation}
The projection condition \eqref{projection} ensures that the weight of any term appearing in  equation \eqref{Killing_Spinor_Simp} is greater than or equal to the sum of weights of the various field components (${\bar{H}}^1_{MNP}$ or $\Psi_M$) that enter that term. For example, for $M=v$ the weight of the equation is 1. Only weight 1 and weight 0 components of ${\bar{H}}_{vNP}^1$ can contribute: in this case there is no other choice, as  weight 1 is the highest weight. However, for $M=u$ the projection condition \eqref{projection} plays a crucial role. Components $\bar{H}_{uNP}^1$ can be of weight $-1$ or of weight $0$. The weight zero term necessarily comes with $\Gamma^v$. However, since $\Gamma^v \epsilon = 0$, and $G^{v P} =0$ for $P \neq u$,  the projection condition ensures that only weight $-1$ fields contribute. 

To see this more explicitly, let us begin by analysing the $u$ equation by setting $M=u$ in equation \eqref{Killing_Spinor_Simp}. 
 For $\epsilon$ independent of $u$, the equation becomes
\be
\quart \omega_{u AB} \widetilde \Gamma^{AB} \epsilon - \quart \bar{H}^1_{uNP} \Gamma^{NP}   \epsilon =0. \label{Killing_Spinor_Simp_u}
\ee 
From the weight argument it follows that both terms $\omega_{u AB} \widetilde \Gamma^{AB} \epsilon $ and $\bar{H}^1_{uNP} \Gamma^{NP}   \epsilon$ are zero, as both metric and the form-field components have at least weight 0. Indeed, the fact that both $\omega_{u AB} \widetilde \Gamma^{AB} \epsilon $ and $\bar{H}^1_{uNP} \Gamma^{NP}   \epsilon$ are zero
 can be checked without much effort. Consider the following vielbeins for the full weight 2 metric, 
\be
\begin{split}
 e^0 &= \psi^{-1}\left[ \half (dv - du) + \zeta \right] \\ 
 \end{split}
 \qquad \; \qquad
 \begin{split}
 e^1 &= dv - e^0,  \\ 
 \end{split}   
\ee
and the $e^2, e^3, e^4, e^5$ as given in equations \eqref{frame2FS_2}--\eqref{frame2FS5}. 
For this choice, the only non-zero components of the spin connection $\o_{uAB}$ turn out to be:
\be
\label{spinGHFS_wt2}
\o_{u03} = \o_{u13} = - \quart r^{1/2} \psi^{-5/2} \psi' .
\ee
Thus for the term $\omega_{u AB} \widetilde \Gamma^{AB} \epsilon$ we have 
\bea
\omega_{u AB} \widetilde \Gamma^{AB} \epsilon  &=& 2\o_{u13} ( \widetilde \Gamma^{13}+  \widetilde \Gamma^{03})\epsilon = 2\o_{u13} \left(e^1_\mu + e^0_\mu\right) e^3_\nu \Gamma^{\mu\nu} \epsilon\\ 
&=& 2 \o_{u03}  e^3_r \Gamma^v \Gamma^r \epsilon = 0,
\eea
where in the last step we have used $\Gamma^v \epsilon = 0$ and $G^{v r} =0$. In Gibbons-Hawking coordinates, the only non-zero component of ${\bar H}^1_{uNP}$ is ${\bar H}^1_{uvr}$. Thus, $ \bar{H}^1_{uNP} \Gamma^{NP} \epsilon= 2 \bar{H}^1_{uvr} \Gamma^{vr}\epsilon  =  0$, since $\Gamma^v \epsilon = 0$, and $G^{v r} =0$. Thus, we conclude that $u$ equation is automatically satisfied, provided the Killing spinors  are $u$ independent and the projection conditions are satisfied.

Let us now set $M=v$ in equation \eqref{Killing_Spinor_Simp}.
The $v$ component of the equation has weight 1. Various terms in the equation can  receive contributions from weight 0 and weight 1 terms in the field configuration.
Demanding Killing spinors to be $v$ independent, the equation becomes,
\be
\quart \omega_{v AB} \widetilde \Gamma^{AB} \epsilon - \quart \bar{H}^1_{vNP} \Gamma^{NP}   \epsilon =0. \label{Killing_Spinor_Simp_v}
\ee 
 In Gibbons-Hawking coordinates, the non-zero components of ${\bar H}^1_{vNP}$ are:
\begin{align}
 {\bar H}^1_{vur} &= - \frac{1}{4} \frac{\psi'}{\psi^2} &
 {\bar H}^1_{vx^4r} &= - \frac{\widetilde{J}}{16 r^2 \psi^2}\\
 {\bar H}^1_{v\phi r} &=  - \frac{\widetilde{J} \cos \theta}{16 r^2 \psi^2}, &
 {\bar H}^1_{v\theta\phi} &= \frac{\widetilde{J} \sin \theta}{16 r \psi}. 
 \end{align}
For weight 0 and weight 1 terms in the metric, a convenient choice for vielbeins is
\be
\begin{split}
 e^0 &= \half(dv - \psi^{-1} (du - 2\zeta)), \\ 
 e^2 &= \psi^{1/2} r^{1/2} (dx^4 + \cos \theta d \phi), \\
 e^4 &= \psi^{1/2} r^{1/2} d \theta,
 \end{split}
 \qquad \; \qquad
 \begin{split}
 e^1 &= \half(dv + \psi^{-1} (du - 2\zeta)),  \\ 
 e^3 &= \psi^{1/2} r^{-1/2} dr, \\
 e^5 &= \psi^{1/2} r^{1/2} \sin \theta d\phi.
 \end{split}   
 \label{framefields_wt_1}
\ee
From this choice, the only non-zero components of the spin connection $\o_{vAB}$ turn out to be:
\be
\begin{aligned}
\label{spinGHFS}
\o_{v03} &= - \o_{v13} = - \quart r^{1/2} \psi^{-3/2} \psi' = + \quart r_0 r^{-3/2} \psi^{-3/2}, \\
\o_{v23} &= - \o_{v45} = - \frac{\widetilde J}{16r^2 \psi^2}.
\end{aligned}
\ee
Using these spin connection coefficients, we see that
\begin{align}
 \o_{v03} \widetilde{\Gamma}^{03} + \o_{v13} \widetilde{\Gamma}^{13} &= \o_{v13} (-\widetilde{\Gamma}^{03} + \widetilde{\Gamma}^{13}) \\
 &= \o_{v13} (-e^0_{\mu} e^3_{\nu} + e^1_{\mu} e^3_{\nu} ) \Gamma^{\mu \nu} \no \\
 &= \o_{v13} \: (-e^0_{\mu} + e^1_{\mu}) \: \psi^{1/2} r^{-1/2} \Gamma^{\mu r} \no \\
 &= \quart \frac{r_0}{r^2 \psi^2} \left(\Gamma^{ur} + \frac{\widetilde J}{4r} \left( \Gamma^{x^4 r} +  \cos \theta \: \Gamma^{\phi r} \right) \right).
\end{align}
Similarly, 
\bea
 \o_{v23} \widetilde{\Gamma}^{23} +   \o_{v45} \widetilde{\Gamma}^{45} &=& \o_{v23} \: e^2_{\mu} e^3_{\nu} \Gamma^{\mu \nu} +  \o_{v45} \: e^4_{\mu} e^5_{\nu} \Gamma^{\mu \nu} \\
& =& - \frac{\widetilde J}{16 r^2 \psi}(\Gamma^{x^4 r} + \cos \theta \: \Gamma^{\phi r}) + \frac{\widetilde J}{16 r \psi} \sin \theta \: \Gamma^{\theta \phi}.
\eea
With these expressions at hand, we readily see that the $v$ equation \eqref{Killing_Spinor_Simp_v} is satisfied
 for all spinors obeying the projection conditions \eqref{projection}--\eqref{projection2}. 
   
Next we analyse $x^4, r, \theta$ and $\phi$ components of the Killing spinor equation. Since these four equations are at weight 0, we only need to consider only weight zero field components. Ignoring weight 1 and 2 terms, we are left with only weight 0 terms in the metric: 
\be
ds^2 =  \psi^{-1}(r) du dv   + \psi(r) ds^2_\rom{flat}.
\ee
We can choose weight 0 vielbeins as,
\begin{align}
\label{frame0FS}
e^0 &= \half (dv - \psi^{-1} du), &
e^1 &= \half (dv + \psi^{-1} du),    \\
e^2 &= \psi^{1/2}(r) \, r^{1/2} (dx^4 +\cos\theta d\phi), &
e^3 &= \psi^{1/2}(r) \, r^{-1/2} \, dr \, , \\
e^4 &= \psi^{1/2}(r)\, r^{1/2}  \, d\theta \, ,  &
e^5 &= \psi^{1/2}(r)\, r^{1/2}  \, \sin\theta\, d\phi\, .
\label{frame0FS3}
\end{align}
The relevant spin connection coefficients turn out to be, 
\begin{align}
&\omega_{x^4}^{23} = \frac{(r \psi)'}{2\psi},& 
&\omega_{x^4}^{45} = \: \half,&  
&\omega_\theta^{25} = \: \half,&  \label{spinw1} \\
&\omega_\theta^{34} = - \frac{(r \psi)'}{2\psi},& 
&\omega_\phi^{23} = \frac{(r \psi)'}{2\psi}\cos \theta, & 
  &\omega_\phi^{24} = - \frac{1}{2} \sin \theta,& \label{spinw2}\\ 
&\omega_\phi^{35} = - \frac{(r \psi)'}{2\psi} \sin \theta,& 
 &\omega_\phi^{45} = - \frac{1}{2} \cos \theta,& 
&\omega_r^{01} = -\frac{\psi'}{2\psi}.& \label{spinw3}
\end{align}
Note that for vielbeins \eqref{frame0FS}
$
e^0 + e^1 = dx^5 + dt = dv.
$
Thus, $\Gamma^v = (e^0_\mu + e^1_\mu)\Gamma^\mu = \widetilde \Gamma^0 + \widetilde \Gamma^1$, 
the projection condition \eqref{projection} can also be written as 
$( \widetilde \Gamma^0 + \widetilde \Gamma^1) \epsilon = 0.
$
It then implies, $
\widetilde \Gamma^0 ( \widetilde \Gamma^0 + \widetilde \Gamma^1) \epsilon = 0,$ i.e., 
\be
\widetilde \Gamma^0  \widetilde \Gamma^1 \epsilon = \epsilon. \label{projection_simp}
\ee
This form of the projection condition proves very convenient.  Let us set  $M=r$ in equation \eqref{Killing_Spinor_Simp}. We have
\be
\left(\partial_r + \frac{1}{2}\omega_{r01} \widetilde \Gamma^{01} - \frac{1}{2} \bar H^1_{ruv} \Gamma^{uv}\right)\epsilon= 0,
\ee
together with $\bar H^1_{ruv} =  \frac{ \psi'}{4\psi^2}.$
Due to the projection condition \eqref{projection_simp} this equation becomes
\be
\left(\partial_r + \frac{\psi'}{4\psi} -   \frac{ \psi'}{8\psi^2} \Gamma^{uv}\right) \epsilon= 0.
\ee
Moreover, we have,  $\Gamma^v = \widetilde \Gamma^0 +  \widetilde \Gamma^1, \Gamma^u = \psi (\widetilde \Gamma^1 -  \widetilde \Gamma^0).
$ Thus, 
\be
\Gamma^{uv} \epsilon = -\frac{1}{2} \Gamma^{v}\Gamma^{u}\epsilon 
= \frac{1}{2} \psi ( \widetilde \Gamma^0 +  \widetilde \Gamma^1)(\widetilde \Gamma^0 -  \widetilde \Gamma^1) \epsilon 
= - 2\psi \epsilon.
\ee 
Hence the radial equation becomes
\be
\left(\partial_r + \frac{\psi'}{2\psi}  \right) \epsilon= 0.
\ee

Since the gravitinos are all of definite chirality, cf.~\eqref{gravchiral}, the supersymmetry variation parameter is also of the same chirality and hence all the Killing spinors we seek are also of the same chirality.  Thus, $\epsilon$ is a six-dimensional left chiral Weyl spinor, i.e.,
\be
\label{KSchiral}
(\widetilde \Gamma_{012345} +1) \epsilon = 0 \implies \widetilde \Gamma^{012345}\epsilon = \epsilon.
\ee
Together with the projection condition \eqref{projection_simp} the chirality condition implies, 
\be
\begin{aligned}
\label{chirality}
\widetilde \Gamma^{23} \epsilon &=& - \widetilde \Gamma^{45} \epsilon, \\
\widetilde \Gamma^{25} \epsilon &=& - \widetilde \Gamma^{34} \epsilon, \\
\widetilde \Gamma^{24} \epsilon &=& + \widetilde \Gamma^{35} \epsilon.
\end{aligned}
\ee

With this input, the $x^4$, $\theta$,  and $\phi$ equations simply become,  
\bea
\partial _{x^4} \epsilon &=& 0, \label{KS_x4}\\
 \partial _\theta \epsilon - \frac{1}{2}  \widetilde \Gamma^{34}  \epsilon &=&0, \label{KS_th}\\
\partial _{\phi} \epsilon - \frac{1}{2}  \sin \theta \: \widetilde \Gamma^{35} \, \epsilon - \frac{1}{2}\cos \theta \: \widetilde \Gamma^{45} \, \epsilon &=&0 \label{KS_phi}.
\eea

To solve for the Killing spinors explicitly, we need to use a representation of $\widetilde \Gamma $ matrices. We use
\begin{align}
\label{gamma01}
\widetilde \Gamma^0 &=   \, \, \,  \mathbb{1}_2 \otimes   \mathbb{1}_2 \otimes (-\mathrm{i}) \sigma_1, &
\widetilde \Gamma^1 &= \, \, \, \mathbb{1}_2 \otimes  \mathbb{1}_2 \otimes \sigma_2, \\
\label{gamma23}
\widetilde \Gamma^2 &=  \, \, \, \mathbb{1}_2\otimes  \sigma_1 \otimes  \sigma_3, &
\widetilde \Gamma^3 &=  \, \, \, \sigma_3 \otimes  \sigma_3 \otimes  \sigma_3, \\
\label{gamma45}
\widetilde \Gamma^4 &=  \, \, \, \sigma_1 \otimes  \sigma_3 \otimes  \sigma_3, &
\widetilde \Gamma^5 &=  \, \, \, \sigma_2 \otimes  \sigma_3 \otimes  \sigma_3.
\end{align}
The above choice has the advantage that $\widetilde \Gamma^3,\widetilde \Gamma^4, \widetilde \Gamma^5$ are represented as three distinct Pauli matrices in the first factor.   Using these matrices we can solve the Killing spinor equations. We find two independent solutions, 
\be
\label{KS_soln_1}
\epsilon = \psi(r)^{-\frac{1}{2}} e^{ \frac{i}{2} \phi} 
\left( 
\begin{array}{c}
\cos \frac{\theta}{2}\\
-\sin \frac{\theta}{2}
\end{array}
\right) \otimes  
 \left( 
\begin{array}{c}
-i\\
1
\end{array}
\right)
 \otimes  \left( 
\begin{array}{c}
1\\
0
\end{array}
\right),
\ee
\be
\label{KS_soln_2}
\epsilon = \psi(r)^{-\frac{1}{2}} e^{- \frac{i}{2} \phi} 
\left( 
\begin{array}{c}
\sin \frac{\theta}{2}\\
\cos \frac{\theta}{2}
\end{array}
\right) \otimes   \left( 
\begin{array}{c}
-i\\
1
\end{array}
\right) \otimes\left( 
\begin{array}{c}
1\\
0
\end{array}
\right).
\ee

Let us count the number of independent Killing spinors. To begin with $\epsilon$ has 32 complex components. There are two conditions for it to be the supersymmetry parameter of (2,0) theory; namely, the chirality condition and  pseudo-Majorana reality condition. This brings the number of independent real spinor components to sixteen.  The projection conditions \eqref{projection}--\eqref{projection2} give the total number of independent Killing spinors for BMPV solution to be 4.

\subsection{Smooth coordinates near the future horizon}
\label{sec:non_singular_coordindate}

In this section, following \cite{Horowitz:1996th, Horowitz:1996cj, Horowitz:1997si, Ross:1997pd, Jatkar:2009yd}, we write the BMPV black string in coordinates such that its metric and form-field are  smooth near the future horizon. These coordinates will then be used to analyse the smoothness of hair modes at the horizon in later sections.  

For simplicity we start with the non-rotating BMPV black string. The metric simplifies to
\be
ds^2 = \psi^{-1}(dudv + K dv^2 ) + \psi\left(r^{-1}
dr^2  + 4\, r \, d\Omega_{3}^2\right) , \label{metric_non_rotating}
\ee
where,
\begin{equation}
K(r) = \psi(r) -1  = \frac{r_0}{r}\, ,
\end{equation}
and $d\Omega_3^2$ given in \eqref{dOmega3}. We do the  following coordinate transformation from $(u,v,r)$ to $(U,V,W)$:
\bea
\label{eq:newcoord1} 
V &=& -\sqrt{r_0}\exp\left(-\frac{v}{\sqrt{r_0}}\right) , \\
\label{eq:newcoord2} 
W&=& \frac{1}{R}\exp \left(\frac{v}{2\sqrt{r_0}}\right)  , \\ 
\label{eq:newcoord3} 
U &=& u + \frac{R^2}{2\sqrt{r_0}} + 2v\, ,
\eea
where
\be 
\label{eq:defR}
R\equiv 2\sqrt{r_0\left(1+ \frac{r_{0}}{r}\right)} \, .
\ee

At the future horizon the standard time coordinate $t$ goes to  infinity, as a result, $v = x^5 + t$ goes to infinity.  As $v \to \infty$, the coordinate $V$ goes to zero from below.  The region outside the horizon has $V<0$.
In reverse, the coordinate transformation \eqref{eq:newcoord1}--\eqref{eq:newcoord3} is,
\bea
\label{eq:newcoord1_rev} 
v &=& \sqrt{r_0} \, \ln \left(- \frac{\sqrt{r_0}}{V}\right), \\
\label{eq:newcoord2_rev} 
r &=& - \frac{4 r_0^{3/2} \, V W^2}{1 + 4 \sqrt{r_0} \, V W^2},\\
\label{eq:newcoord3_rev} 
u &=& U + \frac{1}{2 V W^2} - 2 \sqrt{r_0} \, \ln \left(- \frac{\sqrt{r_0}}{V}\right).
\eea
Inserting  \eqref{eq:newcoord1_rev}--\eqref{eq:newcoord3_rev} in \eqref{metric_non_rotating}, the metric takes the form,
\begin{eqnarray}
ds^2  &=&  4\, r_0\, \bigg[
W^2 dUdV + dV^2 r_0 W^4 Z^{-3} (24+
128\sqrt{r_0}V W^2 + 192 r_0 V^2 W^4)
 \nonumber \\
&& - dVdW  \, 4\sqrt{r_0} W}{Z^{-3}(3+ 12\sqrt{r_0}V W^2
+ 16r_0 V^2
W^4)
  +  W^{-2}Z^{-3}dW^2 + Z^{-1}d\Omega_{3}^2\bigg]\, , \nonumber \\
\end{eqnarray}
where
\be
Z = 1 + 4 \sqrt{r_0} \, V W^2.
\ee
From this expression it is easy to see that the metric is regular (in fact analytic) at the future horizon $V=0$. Near the horizon the metric is locally $\mathrm{AdS}_3 \times \mathrm{S}^3$. We can verify this by computing the Ricci tensor for the above metric. We find that,
\be
R_{MN} + \frac{1}{2r_0} g_{MN} = 0,
\ee
in the $V \to 0$ limit for the $(U, V, W)$ part of the metric;  and 
\be
R_{MN} - \frac{1}{2r_0} g_{MN} = 0,
\ee
in the $V \to 0$ limit for the $x^4, \theta, \phi$ part of the metric. The Ricci tensor and the metric take the block form in the limit $V \to 0$, i.e., there are no cross terms in the  $\mathrm{AdS}_3$ and $\mathrm{S}^3$ parts.

We can also write the three-form field strength in the new coordinates. 
We get
\begin{equation}
  F^{(3)} = \frac{r_0}{\lambda}\, \left[ \sin\theta\,
dx^4\wedge d\theta\wedge d\phi + 4WdW\wedge dV\wedge dU\right]\,  .
\label{three_form_non_rotatingFS}
\end{equation}
$F^{(3)}$ is well behaved and independent
of $V$. One can also easily check the self-duality property of the  $F^{(3)}$. The epsilon convention in the new coordinates become $\epsilon^{U V W x^4 \theta  \phi} = +1.$

For the rotating BMPV black hole,  metric in Gibbons-Hawking coordinates  has the form,
\be
ds^2 = \psi^{-1}\left(dudv + K dv^2 + \frac{\widetilde J}{4r}
\, (dx^4+\cos\theta \: d\phi)
\, dv\right) 
+ \psi\left(r^{-1}
dr^2  + 4\, r \, d\Omega_{3}^2\right) . \label{metric_rotating}
\ee
In order to introduce coordinates in which the metric functions are analytic in the near-horizon region we proceed in three steps. First, we shift $x^4$ coordinate as 
\be
 x^4 = \widetilde x^4 - \frac{\widetilde J}{8r_0^2} v, \label{x4_shift}
\ee
so that the cross term between $d v$ and $(d \widetilde x^4 +\cos\theta \: d\phi) $ has a zero at $r=0$, as in the non-rotating case. The transformed metric takes the form,
\bea
ds^2 &=& \psi^{-1}\left[dudv + \left(K + \frac{\widetilde J^2} {64 r_0^4}
r \psi^2 - \frac{\widetilde J^2}{32 r_0^2 \, r}
\right) dv^2 + \left(\frac{\widetilde J}{4r}
- \frac{\widetilde J r}{4 r_0^2} \psi^2\right)
\, (d\widetilde x^4+\cos\theta \: d\phi)
\, dv\right] \no \\
&& + \psi\left(r^{-1}
dr^2  + 4\, r \, d\widetilde \Omega_{3}^2\right) ,
\eea
where now, 
\be
d\widetilde \Omega_{3}^2 \equiv \quart \left((d\widetilde x^4+\cos\theta \: d\phi)^2
+d\theta^2 +\sin^2\theta \: d\phi^2\right).
\ee
The shift \eqref{x4_shift} changes the identification under $x^5\equiv x^5+2\pi R_5$. The new identification takes the form 
\be
(\widetilde x^4,x^5)\equiv \left(\widetilde x^4 + \frac{\widetilde J}{8 r_0^2} 2\pi R_5, x^5+2\pi R_5\right).
\label{new_identification}
\ee
This, however, does not affect our (local) analysis.

Next, we carry out a rescaling,
\bea
 u &=& \left(1-\frac{\widetilde J^2}{64r_0^3}\right)^{1/2} \widetilde u,\\
 v &=&  \left(1-\frac{\widetilde J^2}{64r_0^3}\right)^{-1/2}\widetilde v,
\eea
so that the coefficient of the $d\widetilde v^2$ term in the metric remains unity as $r\to 0$, as in the non-rotating case. In order to carry out this rescaling we must have, 
\be
\widetilde J^2 <  64r_0^3. \label{J_condition}
\ee
This condition is the cosmic-censorship bound on the angular momentum parameter  of the BMPV black hole. See, e.g., discussion in \cite{Marolf:2005cx}.  Thus, for all  parameter values relevant for the BMPV black hole we can carry out this rescaling. The rescaling gives the metric,
\bea
ds^2 &=& \psi^{-1}\left[d\widetilde u \ d\widetilde v + \left(K +\frac{\widetilde J^2}{64 r_0^4} r \psi^2
-\frac{\widetilde J^2}{32 r_0^2 r}\right)
\left(1-\frac{\widetilde J^2}{64r_0^3}\right)^{-1}d\widetilde v^2\right.
\no \\ 
&& \qquad \left. + \left(\frac{\widetilde J}{4r}
 - \frac{\widetilde J r}{4 r_0^2}\psi^2\right)\,
 \left(1-\frac{\widetilde J^2}{64r_0^3}\right)^{-1/2}
\, (d\widetilde x^4+\cos\theta \: d\phi)
\, d\widetilde v\right] \no \\
&& \qquad + \ \psi\left(r^{-1}
dr^2  + 4\, r \, d\Omega_{3}^2\right) .
\eea

In this metric the coefficient of the $d\widetilde u  d\widetilde v$ term and the coefficients  of the flat space coordinates $(r, \widetilde x^4, \theta, \phi)$ are the same as for the non-rotating BMPV black hole. Moreover, in the $r\to 0$ limit,  the coefficients of the $d\widetilde v^2$ and $(d\widetilde x^4+\cos\theta d\phi)
\, d\widetilde v$ terms have the same numerical values as for the non-rotating BMPV black hole. Thus, as a first guess it is natural to try to the same coordinates as for the non-rotating black hole, i.e., equations \eqref{eq:newcoord1_rev}--\eqref{eq:newcoord3_rev}:
\bea
\label{eq:newcoord1_rev_rotating} 
\widetilde v &=& \sqrt{r_0} \, \ln \left(- \frac{\sqrt{r_0}}{V}\right), \\
\label{eq:newcoord2_rev_rotating} 
r &=& - \frac{4 r_0^{3/2} \, V W^2}{1 + 4 \sqrt{r_0} \, V W^2},\\
\label{eq:newcoord3_rev_rotating} 
\widetilde u &=& U + \frac{1}{2 V W^2} - 2 \sqrt{r_0} \, \ln \left(- \frac{\sqrt{r_0}}{V}\right).
\eea
When we do this transformation, we find that except for the $dV^2$ term all terms are smooth in the $V \to 0$ limit. The $dV^2$ term has a singularity of the form, 
\bea
ds^2 = - \frac{\widetilde J^2}{8 r_0^{3/2}} \left(1- \frac{\widetilde J^2} {64r_0^3}\right)^{-1} W^2 V^{-1} dV^2 + \verb!non-singular terms!
\eea
This singular term, however, can be easily removed by adjusting the coefficient of  the $ \ln \left(- \frac{\sqrt{r_0}}{V}\right)$ term in the $\widetilde u$ transformation \eqref{eq:newcoord3_rev_rotating}. With the transformation, 
\be
\widetilde u = U + \frac{1}{2 V W^2} - \left( 2 \sqrt{r_0} + \frac{\widetilde J^2}{32 r_0^{5/2}} \left(1-\frac{\widetilde J^2}{64r_0^3}\right)^{-1}  \right)  \, \ln \left(- \frac{\sqrt{r_0}}{V} \right).
\label{eq:newcoord3final_rev_rotating} 
\ee
the resulting metric is smooth in the $V \to 0$ limit. The resulting metric is not particularly illuminating, so we do not present  those details (though we use it for our later calculations).

The three-form field strength in the new coordinates is also non-singular. It takes the form, 
\begin{eqnarray}
F^{(3)} &=& \frac{r_0}{\lambda}\, \bigg
[ \sin\theta\, 
d\widetilde x^4\wedge d\theta\wedge d\phi + 4
WdW\wedge dV\wedge  dU   \nonumber \\
&& -  \frac{\widetilde J}{r_0}\, \left(1-\frac{{\widetilde J}^2}
{64 r_0^3}\right)^{-1/2}\,
WdW\wedge dV\wedge
  (d\widetilde x^4 + \cos\theta \:
  d\phi)   \nonumber \\
&&  - \frac{\widetilde J}{2r_0}\left(1-\frac{{\widetilde J}^2}
{64 r_0^3}\right)^{-1/2}W^2 \, \sin\theta\, 
dV\wedge d\theta\wedge d\phi \bigg]\,  .
\label{eq:three_form_rotatingFS}
\end{eqnarray}
It  can be  confirmed that expression \eqref{eq:three_form_rotatingFS} is self-dual.

\section{Deformations of the BMPV black hole}
\label{sec:deformations}

In this section we analyse a class of null deformations of the BMPV black hole generated by the Garfinkle-Vachaspati transform \cite{garfinkle:1990, myers:1997}. The deformations added by this method turn out to be singular \cite{myers:1997, Horowitz:1997si, Jatkar:2009yd} on the BMPV black hole, however, it is instructive to show this in detail as similar modes  turn out to be non-singular for the  BMPV black hole in Taub-NUT. 

\subsection{Bosonic deformations generated by Garfinkle-Vachaspati transform}
\label{sec:bosonic_deformations}

It is useful to identify solution generating techniques that can be exploited to add hair modes on black holes. One such technique is the Garfinkle-Vachaspati transform~\cite{garfinkle:1990, myers:1997}. This method transforms known solutions to new exact solutions of supergravity theory, where the new solution describes a gravitational wave on the original background. The technique works as follows: given a space-time metric $G_{MN}$, we identify a vector field $k_M$ such that it is
\bea
 \text{null} &: \: & k^M k_M = 0, \\
 \text{hypersurface orthogonal} &: \: & \nabla_{[M}k_{N]} = k_{[M}\nabla_{N]} A, \\
 \text{Killing} &: \: &  \nabla_{(M}k_{N)} = 0,
\eea
for some scalar function $A$. New exact solutions to the supergravity equations are constructed by the following transform,
\be
G'_{MN} = G_{MN} + e^A \, T \, k_M \, k_N.
\ee
The above is  a valid solution if $T$ satisfies,
\be
\nabla^2 T = 0  \qquad \: \text{and} \: \qquad k^M \d_M T = 0,
\ee
and matter fields, if present, satisfy certain mild conditions.\footnote{Our configurations satisfy those mild conditions.  It follows from the analysis of reference~\cite{myers:1997} that we do not need to modify the matter fields while deforming the BMPV black hole or the BMPV black hole in Taub-NUT.}

The BMPV black string in six-dimensions possesses such a vector, 
\be 
k^M \partial_M = \frac{\partial}{\partial u}.
\ee 
It is hypersurface orthogonal with $e^A = \psi$. Applying the Garfinkle-Vachaspati transform we get,
\be
ds^2 = \psi^{-1} \left[ du \, dv + (\psi -1 + T(v, \vec w)) \,dv^2 + \chi_i(r) \, dv \, dw^i \right] + \psi \, ds^2_\rom{flat}, \label{deformed_BMPV}
\ee
The six-dimensional Laplacian $\nabla^2$ in the BMPV black hole metric simply reduces to a four-dimensional Laplacian acting on $T(v, \vec w)$: 
\be
\nabla^2 T = \partial_{w^i} \partial_{w^{i}} T = 0.
\ee 
A general solution for $\partial_{w^i} \partial_{w^{i}} T = 0$ can be written as an expansion in spherical harmonics. 

Requiring regularity at infinity and at the origin and keeping only terms that cannot be removed by coordinate transformations \cite{Dabholkar:1995nc}, we can choose
\be
T(v, \vec w) = f_i(v) w^i, \qquad \qquad  \int_0^{2 \pi R_5} f_i(v) dv = 0,
\ee
with four arbitrary functions $f_i(v)$. The deformed metric \eqref{deformed_BMPV} does not look  asymptotically flat, but via a standard change of coordinates \cite{Dabholkar:1995nc} it can be seen to be manifestly asymptotically flat. Further comments on this deformation can be found in \cite{Jatkar:2009yd, Banerjee:2009uk}. We note that the deformation only adds a weight 2 term to the metric and the other fields remain unchanged. As a result, the Killing spinor analysis of section \ref{sec:killing_spinors} remains exactly the same. The deformed solution admits the same Killing spinors \eqref{KS_soln_1}--\eqref{KS_soln_2} as the undeformed solution. Since we are mostly concerned with the question whether the deformation represents a smooth hair mode or not, we next turn to its smoothness analysis. The smoothness analysis below generalises the corresponding discussion of \cite{Jatkar:2009yd} to the rotating BMPV black hole.

The deformation adds the following extra term to the metric,
\be
\delta (ds^2) =  \psi^{-1} T(v, \vec w) dv^2 = \psi^{-1} f_i(v) w^i dv^2 = 2 r^{1/2} \psi^{-1}  f_i(v) m^i dv^2 \label{deform} 
\ee
where $m^i = w^i/|w|$ is the four-dimensional unit vector. The SO(4) unit vector $m^i$ only depends on the angular coordinates. In order for a deformation to be considered as a hair mode, it is necessary that it is only supported outside the horizon. This is most easily analysed in the near-horizon  $\rho, \tau, v$ coordinates introduced in \eqref{NHcoord} with $\beta \rightarrow 0$ limit. In the near-horizon limit,
\be
2 r^{1/2} \, \psi^{-1}(r) \,  f_i(v) \, m^i dv^2 \longrightarrow 2 r_0^{1/2} \, (\beta \, \rho)^{3/2} \, f_i(v) \, m^i dv^2.
\ee
Since these deformations scale as  $\beta^{3/2}$, they vanish in the near-horizon limit. Thus, they are supported outside the horizon and are possible hair modes.

To analyse whether they are smooth or not, we write them in non-singular coordinates  \eqref{eq:newcoord1_rev_rotating}, \eqref{eq:newcoord2_rev_rotating}, and \eqref{eq:newcoord3final_rev_rotating}. The deformation takes the form,
\be
\delta (ds^2) = \frac{1024  }{64 r_0^3-\widetilde{J}^2}  
\frac{r_0^{21/4} W^3}{\sqrt{1+ 4 \sqrt{r_0} W^2 V}} \frac{dV^2}{\sqrt{-V}} f(V, \tilde x^4, \theta, \phi), \label{singular}
\ee
where  $f_i(v) m^i$ is written as $f(V, \tilde x^4, \theta, \phi)$ in the new coordinates. Note that $x^4$ is replaced with $\tilde x^4$; cf.~\eqref{x4_shift}. 
The metric in these coordinates is singular at $V=0$. 

The singular term \eqref{singular} can, however,  be removed by the following shift of the $U$ coordinate, 
\be
U  = \widetilde{U} - G(V, W, \tilde x^4, \theta, \phi),
\ee
with 
\be
 G(V, W, \tilde x^4, \theta, \phi) =  \frac{256 r_0^{17/4}}{64 r_0^3- \widetilde{J}^2} W  \int_0^V \frac{f (V', \tilde x^4, \theta, \phi)}{(1 + 4 \sqrt{r_0} V' W^2)^{1/2}} \frac{dV'}{\sqrt{-V'}} .
\ee
This shift results in a metric that is again smooth near $V=0$, though it  generates additional terms, 
\bea
& & - 4 r_0 W^2 \partial_W  G(V, W, \tilde x^4, \theta, \phi) dW dV - 4 r_0 W^2  \partial_{\theta_i}  G(V, W, \tilde x^4, \theta, \phi)  d \theta^i dV
\eea
where $\theta^i$ collectively denotes $\tilde x^4, \theta, \phi$. These additional terms all vanish in the $V \to 0$ limit. The $V$ derivatives of the function $G$, however, diverge. This hints at possible divergences in the Riemann tensor. Indeed, by an explicit calculation one can check that some components of the Riemann tensor diverge. For example, 
\bea
R_{V W V W} &=& - 2 r_0 W^{-1} \partial_W \partial_V (W^3 \partial_W G) + \verb+non-singular+ \\ 
&=& -\frac{1}{\sqrt{-V}} \frac{1536  r_0^{21/4} }{ 64 r_0^3-\widetilde{J}^2}W f(V, \tilde x^4, \theta, \phi)+ \verb+non-singular+, \eea
and
\bea
R_{V W V \theta^i} &=& - ( 2 r_0 W \partial_W ( W \partial_\theta^i \partial_V G)+ \verb+non-singular+ \\ 
&=&-\frac{1}{\sqrt{-V}} \frac{1024  r_0^{21/4} }{ 64 r_0^3-\widetilde{J}^2} W^2 \partial_\theta^i f(V, \tilde x^4, \theta, \phi)+ \verb+non-singular+ ,
\eea
diverge as $V \to 0$. 

Upon setting $\widetilde{J} =0$ we recover the expressions from appendix C of \cite{Jatkar:2009yd}. 

Since these modes are singular at the horizon, they are not be counted as proper hair modes of the BMPV black hole.

\subsection{Fermionic deformations}
\label{sec:fermionic_deformations}
 
   The $(2,0)$ supergravity theory we are working with has 16 supersymmetries. The  black hole solutions we are working with preserve 4 of these supersymmetries and hence give rise to 12 fermionic zero modes. Out of these 12 zero modes, four are left moving and 8 are right moving. The 4 left moving modes can be elevated to arbitrary functions of $v$ keeping the supersymmetry of the original solution unchanged~\cite{Banerjee:2009uk}.  The aim of this section is to construct these modes and to analyse their smoothness properties.
  
 The linearised equations of motion in the fermionic sector for the gravitino $\psi^{\alpha}_{M}$ were given in section \ref{sec:sugra}, which we rewrite below for convenience: 
\bea
\Gamma^{MNP} D_N \Psi^{\alpha}_P - \bar{H}^{kMNP} \Gamma_N \widehat{\Gamma}^{k}_{\alpha \beta} \Psi^\beta_P &=& 0, \label{gravitino} \\
    H^{sMNP} \Gamma_{MN} \Psi^{\alpha}_P     &=& 0. \label{H-gravitino} 
\eea
Equation \eqref{H-gravitino} is automatically satisfied as all anti-self-dual fields $H^{sMNP} = 0$ for the undeformed background.

We will now solve the gravitino equation \eqref{gravitino} in the undeformed  background and argue that the deformations generated by the Garfinkle-Vachaspati transform do not modify the solutions. We will also argue that the gravitino modes do not back-react, i.e., their stress tensor does not change the background solution.

 We make the following ansatz,
\bea
\Psi_M^\alpha &=& 0 \qquad \text{for} \qquad M \neq v,  \label{ansatz-grav} \\
\partial_u \Psi_M^\alpha &=& 0, \label{u-independence}
\eea
 together with the gauge condition,
\be
\Gamma^M \Psi_M^\alpha = 0. \label{gauge-grav}
\ee
The gauge condition along with the above ansatz reads,
\be
\Gamma^v \Psi_v^\alpha = 0. \label{ansatz+gauge}
\ee

It is evident that ansatz \eqref{ansatz-grav} guarantees that the fermionic deformations being constructed are of weight 1. We now argue that all terms in   equation \eqref{gravitino} must be of weight 1 or more. This is achieved by looking at all the ways of changing the weight of a term and concluding that none of them can decrease the weight of a term involving  $\Psi_v$. There are three potential ways to decrease the weight:

\indent \textendash \: \textit{Multiplying with other background fields:} All fields in the original background, be it the metric or the form field, are of weight $\geq 0$. Thus multiplying the gravitino by these fields can only increase the weight. 

\indent \textendash \: \textit{Acting with $u$-derivatives:} The metric, 3-form field and the gravitino (by ansatz) are all independent of $u$ coordinate. Hence $u$ derivatives cannot reduce the weight. 

\indent \textendash \: \textit{Acting with $\Gamma^v$:}   $\Gamma^v$ annihilates the field $\Psi_v^\alpha$, cf.~\eqref{ansatz+gauge}. Hence terms of the form $\Gamma^v \Psi_v$ cannot reduce the weight of a term. In expanding the antisymmetric $\Gamma^{MNP}$, we may find other gamma matrices sandwiched between $\Gamma^v$ and $\Psi_v^\alpha$. Such a  $\Gamma^v$ can be shifted next to the gravitino. The additional terms obtained by the use of the Clifford identity do not decrease the weight. For instance, we might have a term $\Gamma^v \Gamma^M \Psi_v= -\Gamma^M \Gamma^v \Psi_v + 2G^{vM} \Psi_v = 2G^{vM} \Psi_v$. The inverse metric is such that $G^{vM} \neq 0$ only for $M=u$ and this does not decrease the weight.

In then follows that the choice  $M=u$, $N=i$, and $P = v$ in \eqref{gravitino}  gives the only non-trivial equation,
 \be
 \Gamma^{uiv} \Big( \partial_i + \quart {\omega_i}_{AB} \widetilde{\Gamma}^{AB} \Big) \Psi^{\alpha}_v - \bar{H}^{1 uiv} \Gamma_i \widehat{\Gamma}^{1}_{\alpha \beta} \Psi^\beta_v = 0.
 \ee
 This is a weight 1 equation.  
 
The contravariant $u$ index carries with it the total weight of the equation and hence terms in the background fields with weight $> 0$ do not contribute. Thus, the weight 0 vielbeins \eqref{frame0FS}--\eqref{frame0FS3} and the corresponding spin connection coefficients \eqref{spinw1}--\eqref{spinw3} can be used. Furthermore, the form field can be truncated to $F^{(3)}_0 = \frac{r_0}{\lambda} (\eps_3 + \star_6\eps_3)$ to only capture its weight 0 components.

Using Clifford identity, we see that
\be
\Gamma^{uiv} = - G^{vi} \Gamma^u + G^{uv} \Gamma^i,
\ee
in which the truncated metric ensures the presence of the second term alone. Similarly,
\bea
\bar{H}^{1 uiv} \Gamma_i &=& G^{uN} G^{vP} \bar{H}^1_{NiP}\Gamma^i, \\
                         &=& G^{uv} G^{vu} \bar{H}^1_{viu}\Gamma^i.
\eea
 Choosing the transverse coordinates to be the Gibbons-Hawking coordinates $i=(r, \theta, \phi, x^4)$, and dropping an overall factor of $ G^{uv} = 2\psi$, we set out to expand the following equation:
\be
\label{grav-wt1}
\Gamma^i \Big( \partial_i + \quart {\omega_i}_{AB} \widetilde{\Gamma}^{AB} \Big) \Psi^{\alpha}_v + \Gamma^i G^{vu} \bar{H}_{1 ivu}  \widehat{\Gamma}^{1}_{\alpha \beta} \Psi^\beta_v = 0.
\ee
The only non-zero form field components are $\bar{H}^{1 uiv} = \bar{H}^{1 urv} = \frac{\psi'}{4 \psi^2}.$ In section \ref{sec:killing_spinors} it was noted that our choice of vielbeins implies $\Gamma^v = \widetilde{\Gamma}^0 + \widetilde{\Gamma}^1$ and this translates the gauge condition \eqref{ansatz+gauge} to
\be
\label{Gamma01_Psi}
\widetilde{\Gamma}^0 \widetilde{\Gamma}^1 \Psi_v^\alpha = \widetilde{\Gamma}^{01} \Psi_v^\alpha = \Psi_v^\alpha.
\ee

We separately look at the terms corresponding to each of the transverse coordinates starting with $i=r$:
\be
\Gamma^r \Big( \partial_r + \half {\omega_r}_{01} \widetilde{\Gamma}^{01} \Big) \Psi^{\alpha}_v + \Gamma^r G^{vu} \bar{H}_{1 rvu}  \widehat{\Gamma}^{1}_{\alpha \beta} \Psi^\beta_v.
\ee
Since $\widetilde{\Gamma}^3 = e^3_M \Gamma^M = e^3_r \Gamma^r$ we have $\Gamma^r = r^{1/2} \psi^{-1/2} \widetilde{\Gamma}^3$. With these, we obtain,
\be
r^{1/2} \psi^{-1/2} \widetilde{\Gamma}^3 \Bigg( \partial_r + \quart \frac{\psi'}{\psi} - \half \frac{\psi'}{\psi} \widehat{\Gamma}^1 \Bigg) \Psi_v.
\ee

The gravitino $\Psi_v^\alpha$ being a six-dimensional left chiral Weyl spinor satisfies the following chirality conditions, cf.~\eqref{chirality},
\bea
\widetilde \Gamma^{23} \Psi_v &=& - \widetilde \Gamma^{45} \Psi_v, \\
\widetilde \Gamma^{25} \Psi_v &=& - \widetilde \Gamma^{34} \Psi_v, \\
\widetilde \Gamma^{24} \Psi_v &=& + \widetilde \Gamma^{35} \Psi_v.
\eea
These chirality conditions are used as and when needed. The terms contributing to $i = x^4$ are
\be
\Gamma^{x^4} \Bigg( \partial_{x^4} + \half \omega_{x^4 23} \widetilde{\Gamma}^{23} + \half \omega_{x^4 45} \widetilde{\Gamma}^{45} \Bigg) \Psi_v.
\ee
Replacing $\Gamma^{x^4} = (r \psi)^{-1/2} (\widetilde{\Gamma}^2 - \cot\theta \, \widetilde{\Gamma}^5)$ and simplifying the spin connection terms we obtain,
\be
(r \psi)^{-1/2} (\widetilde{\Gamma}^2 - \cot\theta \, \widetilde{\Gamma}^5) \partial_{x^4} \Psi_v + r^{1/2} \psi^{-1/2} \Bigg( \widetilde{\Gamma}^3 \frac{\psi'}{4\psi} - \widetilde{\Gamma}^4 \cot\theta \frac{\psi'}{4\psi} \Bigg) \Psi_v.
\ee

The $i= \theta$ terms give,
\be
\Gamma^{\theta} \Bigg( \partial_{\theta} + \half \omega_{\theta 25} \widetilde{\Gamma}^{25} + \half \omega_{\theta 34} \widetilde{\Gamma}^{34} \Bigg) \Psi_v.
\ee
The relation $\Gamma^{\theta} = (r \psi)^{-1/2} \widetilde{\Gamma}^4$ helps to bring the above terms to the form,
\be
(r \psi)^{-1/2} \widetilde{\Gamma}^4 \partial_{\theta} \Psi_v + r^{1/2} \psi^{-1/2} \widetilde{\Gamma}^3 \Bigg( \frac{1}{2r} + \frac{\psi'}{4 \psi} \Bigg) \Psi_v.
\ee

Finally, contributions from $i = \phi$ give a derivative term and the four spin connection coefficients $\omega_{\phi 23}$, $\omega_{\phi 24}$, $\omega_{\phi 35}$ and $\omega_{\phi 45}$. We make use of the relation $\Gamma^{\phi} = (r \psi)^{-1/2} ({\sin\theta})^{-1} \widetilde{\Gamma}^5$ to get
\be
\Bigg\{ \Big((r \psi)^{-1/2} {\sin\theta}^{-1} \widetilde{\Gamma}^5 \Big) \partial_{\phi} + r^{1/2} \psi^{-1/2} \Bigg( \widetilde{\Gamma}^4 \cot\theta \frac{\psi'}{4 \psi} + \widetilde{\Gamma}^3 \frac{1}{2r} + \widetilde{\Gamma}^3 \frac{\psi'}{4 \psi} \Bigg) + \half (r \psi)^{-1/2} \cot\theta \, \widetilde{\Gamma}^4 \Bigg\} \Psi_v.
\ee

Putting together all the above contributions, we end up with an equation that we want to solve,
\bea
&&r^{1/2} \psi^{-1/2}\, \widetilde{\Gamma}^3 \Bigg( \partial_r +\frac{\psi'}{\psi} + \frac{1}{r} - \half \frac{\psi'}{\psi} \widehat{\Gamma}^1 \Bigg) \Psi_v + \Big((r \psi)^{-1/2} ({\sin\theta})^{-1} \widetilde{\Gamma}^5 \Big) \partial_{\phi} \Psi_v \no \\ 
&&+ (r \psi)^{-1/2} \, (\widetilde{\Gamma}^2 - \cot\theta \, \widetilde{\Gamma}^5) \partial_{x^4} \Psi_v + (r \psi)^{-1/2} \widetilde{\Gamma}^4 \Bigg( \partial_{\theta} + \half \cot\theta \Bigg) \Psi_v = 0.
\label{grav eqn}
\eea
Solutions to this equation were qualitatively predicted in \cite{Banerjee:2009uk} from the zero-mode considerations. The solutions should have no momentum along the $x^4$ direction and $m=\pm 1/2$ units of momentum along the $\phi$  direction. Incorporating these eigenvalues, the partial differential equation simplifies to,
\bea
&&r^{1/2}\psi^{-1/2} \widetilde{\Gamma}^3 \Bigg( \partial_r +\frac{\psi'}{\psi}  - \half \frac{\psi'}{\psi} \widehat{\Gamma}^1 \Bigg) \Psi_v   \no \\
&& + \: (r \psi)^{-1/2} \widetilde{\Gamma}^4 \Bigg[ im {(\sin\theta)}^{-1} \widetilde{\Gamma}^4 \widetilde{\Gamma}^5  +  \widetilde{\Gamma}^4  \widetilde{\Gamma}^3  + \Bigg( \partial_{\theta} + \half \cot\theta \Bigg) \Bigg] \Psi_v = 0.
\eea
 
This form enables us to achieve a separation of variables. We choose a convenient gamma matrix representation indicated in \eqref{gamma01}--\eqref{gamma45}.
Choosing either projection  condition ${\widehat{\Gamma}}^1 \Psi_v = \pm \Psi_v$, the equation involving $r$ (with a zero separation constant) becomes,
\be
\Bigg( \partial_r +\frac{\psi'}{\psi}  \mp \half \frac{\psi'}{\psi} \Bigg) R(r) = 0,
\ee
with solutions, $R(r) = \psi(r)^{-1/2}$ or $R(r) = \psi(r)^{-3/2}$. Thus, we have 
\bea
\Psi_v &=& \psi^{-3/2} \eta(v,\theta, \phi) \qquad \text{for}\qquad {\widehat{\Gamma}}^1 \eta = -\eta,  \label{first_sol} \\
\Psi_v &=& \psi^{-1/2} \eta(v,\theta, \phi) \qquad \text{for}\qquad {\widehat{\Gamma}}^1 \eta = \eta. \label{second_sol} 
\eea

The equation that fixes the $\theta$-dependence becomes,
\be
\Bigg[\partial_{\theta} + \half \cot\theta - m {(\sin\theta)}^{-1} \sigma^3  - i \sigma^2 \Bigg] \eta(v,\theta, \phi) = 0.
\ee
A simple calculation tells us that the two possible solutions are
\begin{align}
\eta(v,\theta, \phi) &= h(v) \: e^{i \phi/2}
\begin{pmatrix}
\cos{(\theta/2)} \\ 
-\sin{(\theta/2)}
\end{pmatrix}
\qquad \text{for} \: \:m = 1/2,  \\
\eta(v,\theta, \phi) &= h(v) \: e^{-i \phi/2}
\begin{pmatrix}
\sin{(\theta/2)} \\
\cos{(\theta/2)}
\end{pmatrix}
\qquad \text{for} \: \:m = -1/2. 
\end{align}
Having no constraint imposed on the $v$ dependence, $h(v)$ is an arbitrary periodic function of the $v$ coordinate. 
The spinorial properties of $\Psi_v$ are completely captured by $\eta$ making it both an SO(5,1) spinor as well as an SO(5) spinor.

How do these solutions behave in non-singular coordinates  given in \eqref{eq:newcoord1_rev_rotating}, \eqref{eq:newcoord2_rev_rotating}, and \eqref{eq:newcoord3final_rev_rotating}?  The gravitino
  configuration with ${\widehat{\Gamma}}^1 \eta = \eta$ behaves as,
 \be
 \Psi_V = \psi^{-1/2} \eta(v,\theta,\phi) = \left(  \frac{\partial v}{\partial V} \right) \Psi_v = \frac{16 r_0^{9/4}}{ (64 r_0^3-\widetilde{J}^2)^{1/2}} \, \frac{1}{\sqrt{-V}} \, W \, \eta(v(V), \theta, \phi)
 \ee
 and the gravitino  configuration with ${\widehat{\Gamma}}^1 \eta = -\eta$ behaves as,
 \be
\Psi_V = \psi^{-3/2} \eta(v,\theta,\phi) = \left(  \frac{\partial v}{\partial V} \right) \Psi_v =  \frac{64 r_0^{11/4}}{ (64 r_0^3-\widetilde{J}^2)^{1/2}} \, \sqrt{-V} \, W^3 \, \eta(v(V), \theta, \phi).
\ee
However, we cannot comment the smoothness of the gravitino field by looking at these expressions. The gravitino field  was computed using  vielbeins \eqref{frame2FS}--\eqref{frame2FS5}.  These vielbeins are singular in the non-singular coordinates.

To see this, let us write vielbeins \eqref{frame2FS}--\eqref{frame2FS5} in the new coordinates. They take the form, 
\bea
e^+ &:=& e^0 + e^1 = - \frac{8r_0^2}{(64 r_0^3-\widetilde{J}^2)^{1/2}} \frac{dV}{V},\\
e^- &:=& \no e^1 - e^0 =\frac{1}{(64 r_0^3-\widetilde{J}^2)^{1/2}}\left( r_0^{-1/2} \widetilde{J}^2 W^2- 96 r_0^{5/2}   W^2  +\frac{8 r_0^2}{V} \right) dV \\  
&& 
+ \, \frac{(64 r_0^3-\widetilde{J}^2)^{1/2}}{2r_0} \left(- V W^2 dU + \frac{dW}{W} \right) \no \\
&& 
+ \,  \widetilde{J}  (4 r_0)^{-1} (1+ 4 \sqrt{r_0} V W^2)(d \tilde x^4 + \cos \theta \, d \phi),  
\eea
and
\bea
e^2 &=&  \frac{\sqrt{r_0}}{(1+ 4 \sqrt{r_0} V W^2)^{1/2}} \left( d \tilde x^4 + \cos \theta \, d \phi  + \frac{\widetilde{J}}{(64 r_0^3-\widetilde{J}^2)^{1/2}} \frac{dV}{V} \right), \\
e^3 &=&  \frac{2 \sqrt{r_0}}{(1+ 4 \sqrt{r_0} V W^2)^{3/2}} \left( \frac{dW}{W} + \frac{dV}{2V}\right) ,\\
e^4 &=& \frac{\sqrt{r_0}}{(1+ 4 \sqrt{r_0} V W^2)^{1/2}} d\theta,\\
e^5 &=& \frac{\sqrt{r_0}}{(1+ 4 \sqrt{r_0} V W^2)^{1/2}} \sin \theta \, d \phi.
\eea
Note that $e^+, e^-, e^2, e^3$ are singular at $V=0$. The metric in non-singular coordinates can be expressed as
\be
ds^2 = e^+ e^- + (e^2)^2 +(e^3)^2 + (e^4)^2 + (e^5)^2  .
\ee
A non-singular set of vielbeins can be obtained by a sequence of Lorentz transformations: first, 
\begin{align}
& \hat e^+ = \alpha e^+, &
& \hat e^- = \frac{1}{\alpha} e^-, 
& \hat e^2 = e^2, &
& \hat e^3 = e^3,
\end{align}
then, 
\begin{align}
& \check e^+ = \hat e^+,&
& \check e^- = \hat e^- - 2 \beta \hat  e^3 - \beta^2 \hat e^+,
& \check e^2 = \hat e^2, &
& \check e^3 = \hat  e^3 + \beta \hat e^+,
\end{align}
and finally, 
\begin{align}
& \tilde e^+= \check e^+ ,&
& \tilde e^- = \check e^- - 2 \gamma \check e^2 - \gamma^2 \check e^+,
& \tilde e^2 = \check e^2 + \gamma \check e^+ ,&
& \tilde e^3 = \check e^3,
\end{align}
where
\bea
\alpha&=&- \frac{(64 r_0^3-\widetilde{J}^2)^{1/2}}{4 r_0^{3/2}} V, \\
\beta&=&-\frac{1}{2 V (1+ 4 \sqrt{r_0} V W^2)^{3/2}}, \\
\gamma &=& -\frac{\widetilde{J}}{2 V (1+ 4 \sqrt{r_0} V W^2)^{1/2} (64 r_0^3-\widetilde{J}^2)^{1/2}}.
\eea
The new vielbeins $\tilde e^+, \tilde e^-, \tilde e^2, \tilde e^3$ are all regular and a direct calculation shows that in the non-singular coordinates
\be
ds^2 = \tilde e^+\tilde e^- + (\tilde e^2)^2 +(\tilde e^3)^2 + (\tilde e^4)^2 + (\tilde e^5)^2.
\ee
The $\beta$ and $\gamma$ transformations are examples of \emph{null rotations} of $e^-$ about $e^+$~\cite{Ortaggio:2012jd}.

 These local Lorentz transformations act on the gravitino field. As shown in appendix \ref{app:Lorentz} the combined action is simply
 \be
 \widetilde{\Psi}_V = \widehat{\Psi}_V = \frac{1}{\sqrt{\alpha}} \Psi_V.
 \ee
For the gravitino configuration with ${\widehat{\Gamma}}^1 \eta = -\eta$, we have
\be
 \widetilde{\Psi}_V = \frac{128 r_0^{7/2}}{ (64 r_0^3-\widetilde{J}^2)^{3/4}}  \, W^3 \, \eta(v(V), \theta, \phi).
 \ee
 This field is well behaved at the horizon, though it does not vanish at the horizon.  We can make it vanish by doing a local supersymmetry transformation with a parameter proportional to 
 \be
  W^3 \,  \int^V_0  \eta(v(V'), \theta, \phi) dV'.
 \ee
 
 The second solution to the fermionic deformation equation with ${\widehat{\Gamma}}^1 \eta = \eta$, cf.~\eqref{second_sol}, is singular, 
 \be
 \widetilde{\Psi}_V = \widehat{\Psi}_V = \frac{1}{\sqrt{\alpha}} \Psi_V = -\frac{32 r_0^{3}W}{ (64 r_0^3-\widetilde{J}^2)^{3/4}} \, \frac{1}{V} \, \eta(v(V), \theta, \phi).
 \ee
 Hence, it is not an allowed deformation.

 In the new coordinate system and the new Lorentz frame, the Killing spinors behaves as 
 \be
 \frac{4 r_0}{ (64 r_0^3-\widetilde{J}^2)^{1/4}}  \, W,
 \ee
 and hence are well defined at the horizon. 
 
Let us now count the independent left-moving smooth gravitino modes. To begin with $\Psi^\alpha_v$ is a 32 component complex spinor. The chirality condition ensures that only 16 of the 32 components are independent. The symplectic Majorana reality condition reduces the number to 16 real components. The gauge condition ${\widetilde{\Gamma}}^{01} \eta = \eta$ and the eigenvalue equation ${\widehat{\Gamma}}^1 \eta = \pm \eta$ brings it down to 8 components (4 with ${\widehat{\Gamma}}^1 \eta = \eta$ and 4 with ${\widehat{\Gamma}}^1 \eta = -\eta$).  We saw that the ${\widehat{\Gamma}}^1 \eta = \eta$ solutions  are singular at the horizon. Hence, we have only four independent  left-moving gravitino modes \cite{Banerjee:2009uk, Jatkar:2009yd}. 

These modes  go to zero in the near-horizon limit as $\beta^{3/2}$, and hence are genuine hair modes.

It may seem that since the above solutions are obtained using the linearised gravitino equation, they may not be solutions when possible  non-linear terms are included in the gravitino equation. This is not the case. These  solutions remain solutions even after taking into account possible non-linearities. This is because, the gravitino equation will remain a weight 1 equation.  The non-linear terms  will  necessarily of be weight 2 or more and hence they will not contribute to the weight 1 equation.

The deformation generated by the Garfinkle-Vachaspati transform also does not change the gravitino solutions. This also follows from weight considerations. The minimum weight at which the Garfinkle-Vachaspati deformation term can contribute in the gravitino equation is 3, since the deformation itself is of weight 2 and  gravitino is of weight 1. However, since the gravitino equation is of weight 1, such terms  cannot contribute to the gravitino equation.

Let us now address the back-reaction that the gravitino deformations can produce.  The dilaton equation is weight zero and hence is completely unaffected by  the gravitino deformations.  The $vv$ component of Einstein's equation prima facie can get a contribution from gravitino bilinears constructed with even number of gamma matrices. However, all such weight 2 terms vanish by the combination of the ansatz \eqref{ansatz+gauge} and properties of the spinor fields in six-dimensions. 

It remains to comment on the supersymmetry of the fermionic deformation  modes. These modes affect two of the Killing spinor equations \eqref{del_metric} and \eqref{del_FF}. Consider equation \eqref{del_metric} with the insertion of an identity matrix: $(\bar \epsilon \, {\widehat{\Gamma}}^1) \, \widetilde \Gamma^A \, ({\widehat{\Gamma}}^1 \, \Psi_M)$. The projection condition on $\eps$ \eqref{projection2} implies that $\bar{\eps} \, {\widehat{\Gamma}}^1 = \bar{\eps}$. The smoothness analysis picks out the solution ${\widehat \Gamma}^1 \, \Psi_M = -\Psi_M$  for the gravitino hair mode. These opposite SO$(5)$ projection conditions guarantee that equation \eqref{del_metric} is satisfied. We note here that the gravitino solutions with ${\widehat \Gamma}^1 \, \Psi_M = \Psi_M$ are not only singular at the future horizon, but also break supersymmetry, as equation \eqref{del_metric} cannot be satisfied.

The fermionic deformations give the weight 1 term $
\bar \eps \, \Gamma_{w^i} \, {\widehat \Gamma}^i \, \Psi_{v}$ for equation \eqref{del_FF}. The SO(5,1) chirality and projection conditions imply $\widetilde \Gamma^{2345} \Psi_v = \Psi_v$ and $\widetilde \Gamma^{2345} \eps = \eps$.  Inserting $(\widetilde \Gamma^2 \widetilde \Gamma^3 \widetilde \Gamma^4 \widetilde \Gamma^5)^2 = 1$ as 
\be
\bar \eps \, \Gamma_{w^i} \, {\widehat \Gamma}^i \, (\widetilde \Gamma^2 \widetilde \Gamma^3 \widetilde \Gamma^4 \widetilde \Gamma^5)^2 \Psi_{v} 
\ee
and moving $\widetilde \Gamma$'s appropriately we see that contributions to \eqref{del_FF} vanish. Hence, the fermionic hair modes preserve the supersymmetry of the original background.

\section{BMPV black hole in Taub-NUT space}
\label{sec:black_hole_BMPV_TN}
In this section, we review the BMPV black hole in Taub-NUT space \cite{Gauntlett:2002nw, hep-th/0503217}. In section \ref{sec:metric_TN}  coordinates, metric, three-form field strength, and near-horizon geometry are presented. In section \ref{sec:killing_spinors_TN} Killing spinors are constructed. In section \ref{sec:non_singular_coordindate_TN}  a set of coordinates is presented  in which the black hole metric is smooth at the future horizon.

\subsection{Metric and form field}
\label{sec:metric_TN}
The four dimensional Taub-NUT space in Gibbons-Hawking coordinates is given by
\be
ds^2_\rom{TN} = \left( \frac{4}{R_4^2} + \frac{1}{r} \right)^{-1} (dx^4 + \cos \theta d\phi)^2 + \left(\frac{4}{R_4^2} + \frac{1}{r}\right) (dr^2 + r^2 d \theta^2 + r^2 \sin^2 \theta d\phi^2).
\ee
Compared to flat space in Gibbons-Hawking coordinates, the only difference is that $\frac{1}{r}$ factors in flat space metric are replaced with $\left( \frac{4}{R_4^2} + \frac{1}{r} \right)$.
The $x^4$ coordinate labels the circle ${\widetilde{\mathrm{S}}}^1$ and it is periodic with size $2 \pi R_4$. Introducing the one form 
\be
\widetilde{\z} = -\frac{\widetilde J}{8} \left(  \frac{4}{R_4^2} + \frac{1}{r}  \right) (dx^4 + \cos \theta d \phi),
\ee
the six-dimensional metric of the BMPV black hole in Taub-NUT space takes the form, 
\be
\label{metric_TN}
ds^2 = \psi^{-1}(r) [du dv + (\psi(r) -1) dv^2 - 2 \widetilde{\z} dv] + \psi(r) ds^2_\rom{TN}.  
\ee
The field strength $F^{(3)}$ supporting this solution is self-dual,
\be
 F^{(3)} = \frac{r_0}{\lambda}\Big[ (\e_3 + \ast_6 \e_3) + \frac{1}{r_0} \psi^{-1}(r) dv \wedge d\widetilde{\z} \Big].
 \label{F3_TN}
\ee
As with the BMPV black hole, the dilaton is set to its asymptotic value $e^{\Phi} = \lambda$.

Recall the set of coordinates $(\r, \t)$ introduced in \eqref{NHcoord} for obtaining the near-horizon geometry by taking the  $\beta \to 0$ limit. The near-horizon limit for the BMPV black hole in Taub-NUT space  coincides with that obtained for the BMPV black hole in flat space \eqref{NHline}. The form field strength $F^{(3)}$ in the near-horizon limit also matches with \eqref{F3_NH}. 

A set of vielbeins can be introduced as was done in \eqref{frame2FS}--\eqref{frame2FS5} with $\z$ replaced with $\widetilde{\z}$ and with appropriate factors of 
\be
\label{TN function}
\chi(r) = \left( \frac{4}{R_4^2} + \frac{1}{r} \right)
\ee
inserted. They are
\bea
\label{frame2FS_TN}
e^0 &=& \psi^{-1}(r) (dt+ \widetilde{\z}), \\
e^1 &=& \left(dv- \psi^{-1}(r) (dt+ \widetilde{\z}) \right), \\
e^2 &=& \psi^{1/2} \, \chi^{-1/2} \, (dx^4 +\cos\theta \, d\phi), \\
e^3 &=& \psi^{1/2}(r) \, \chi^{1/2} \,  dr \, ,  \\
e^4 &=& \psi^{1/2}(r) \, \chi^{1/2} \, r \, d\theta \, ,  \\
e^5 &=& \psi^{1/2}(r) \, \chi^{1/2} \, r \, \sin\theta\, d\phi\, .
\label{frame2FS5_TN}
\eea
Metric \eqref{metric_TN} and three-form field \eqref{F3_TN} can accordingly be expressed in terms of vielbeins \eqref{frame2FS_TN}--\eqref{frame2FS5_TN} analogous to equations \eqref{metric_frame} and \eqref{form_field_frame}.

\subsection{Killing spinors}
\label{sec:killing_spinors_TN}

The following construction of Killing spinors closely parallels the discussion in section \ref{sec:killing_spinors}.

We again demand the projection conditions $\G^v \e = ({\widetilde \G}^0 + {\widetilde \G}^1) \e = 0$ and ${\widehat{\G}}^1 \e = \e$. These conditions simplify the Killing spinor equation to equation \eqref{Killing_Spinor_Simp}. We analyse this equation for different values of $M$. 

Setting $M = v$ we get a weight 1 equation. Taking $\e$ to be  $v$ independent, we get
\be
\quart \o_{vAB} {\widetilde \G}^{AB} \e - \quart {\bar H}^1_{vNP} \G^{NP} \e = 0.
\label{v_eq_KS_TN}
\ee
This equation  only receives contributions  from  weight 0 and weight 1 fields.  A set of vielbeins for the simplified metric with only weight 0 and weight 1 terms takes the form, 
\begin{align}
\label{frameTN01}
& e^0 = \half (dv - \psi^{-1} (du - 2\widetilde{\z})), &
& e^1 = \half (dv + \psi^{-1} (du - 2\widetilde{\z})), \\
\label{frameTN23}
& e^2 = \psi^{1/2} \chi^{-1/2} \: (dx^4 + \cos \theta \, d\phi), &
& e^3 = \psi^{1/2} \chi^{1/2} \: dr, \\
\label{frameTN45}
& e^4 = \psi^{1/2} \chi^{1/2} \: r d\theta, &
& e^5 = \psi^{1/2} \chi^{1/2} \: r \sin \theta \, d\phi.
\end{align}
As in the case of flat space, the above choice of vielbeins is such that $e^0 + e^1 = dv$. We also note that $e^0 -e^1 = \psi^{-1}(du - 2\widetilde \z)$.   Using these vielbeins, we find the spin connection coefficients  $\o_{M}{}^{AB}$. The non-zero $\o_{vAB}$  components are
\begin{align}
\label{spinvTN}
 \o_{v03} &= - \: \o_{v13} = \quart \frac{\psi'}{\psi^{3/2}} \chi^{-1/2},& \o_{v23} &= - \: \o_{v45} = - \frac{\widetilde J}{16 r^2 \psi^2}.
\end{align}
From these expressions, we have,
\be
 \o_{v03} \widetilde{\Gamma}^{03} + \o_{v13} \widetilde{\Gamma}^{13} 
 = \quart \frac{r_0}{r^2 \psi^2} \Big[ \Gamma^{ur} + \frac{\widetilde J}{4} \chi ( \Gamma^{x^4 r} + \cos \theta \: \Gamma^{\phi r}) \Big],
\ee
and
\be
 \o_{v23} {\widetilde \G}^{23} + \o_{v45} {\widetilde \G}^{45}  
 = - \frac{\widetilde J}{16 r^2 \psi} \: (\Gamma^{x^4 r} + \cos \theta \: \Gamma^{\phi r}) + \frac{\widetilde J \sin \theta}{16 \psi} \G^{\theta \phi} \chi.
\ee
The contributing field strength components ${\bar H}^1_{vNP}$ are,
\be
\begin{split}
 {\bar H}^1_{vur} &= - \quart \frac{\psi'}{\psi^2}, \\
 {\bar H}^1_{v \theta \phi} &= \frac{\widetilde J \sin \theta}{16 \psi} \chi,
 \end{split}
 \qquad \: \:
 \begin{split}
 {\bar H}^1_{vr \phi} &= \frac{\widetilde J \cos \theta}{16 r^2 \psi} + \frac{\widetilde J \cos \theta}{16} \frac{\psi'}{\psi^2} \chi, \\
 {\bar H}^1_{vrx^4} &= \frac{\widetilde J}{16 r^2 \psi} + \frac{\widetilde J}{16} \frac{\psi'}{\psi^2}  \chi.
 \end{split}
\ee
Plugging in these expressions, we see that the $v$ equation, cf. 
\eqref{v_eq_KS_TN},  is satisfied. 

Setting $M = u$ in \eqref{Killing_Spinor_Simp} we get, 
\be
\left( \d_u + \half (\o_{u03} {\widetilde{\G}}^{03} + \o_{u13} {\widetilde{\G}}^{13}) - \half {\bar H}^1_{uvr} \G^{vr} \right) \e = 0.  \label{u_eq_KS_TN}
\ee
The relevant $\o_{uAB}$ coefficients are
$\o_{u03} = \o_{u13} = - \quart \frac{\psi'}{\psi^{5/2}} \chi^{-1/2}.
$ This implies that 
$
 \o_{u03} {\widetilde{\G}}^{03} + \o_{u13} {\widetilde{\G}}^{13}  = - \quart \frac{\psi'}{\psi^2} \G^{vr}.
$
Using $\G^{vr} = - \G^r \G^v$ and taking the Killing spinor to be $u$ independent we see that equation \eqref{u_eq_KS_TN} is satisfied.

The  other four equations are all of weight 0.  A convenient choice of weight 0 truncated vielbeins is,
\begin{align}
\label{frame0TN01}
& e^0 = \half (dv - \psi^{-1} du), &
& e^1 = \half (dv + \psi^{-1} du), \\
\label{frame0TN23}
& e^2 = \psi^{1/2} \chi^{-1/2} \: (dx^4 + \cos \theta d\phi), &
& e^3 = \psi^{1/2} \chi^{1/2} \: dr, \\
\label{frame0TN45}
& e^4 = \psi^{1/2} \chi^{1/2} \: r d\theta, &
& e^5 = \psi^{1/2} \chi^{1/2} \: r \sin \theta d\phi.
\end{align}
The spin connection coefficients  $\o_i{}^{AB}$ that arise from these are
\begin{align}
	\label{spin0TNr}
& \o_{r01} = \frac{\psi'}{2 \psi} & 
& \o_{\theta 25} = \frac{1}{2r} \chi^{-1} \\
& \o_{\theta 34} = - \Big( \frac{4}{R_4^2} + \frac{1}{2r} \Big) \chi^{-1} - \frac{r \psi'}{2 \psi} &  
&\o_{x^4 23} = \frac{1}{2 r^2} \chi^{-2} + \frac{\psi'}{2 \psi} \chi^{-1} \\  
&\o_{x^4 45} = \frac{1}{2 r^2} \chi^{-2} &
&\o_{\phi 23} = \Bigg( \frac{1}{2r^2} \chi^{-2} + \frac{\psi'}{2 \psi} \chi^{-1} \Bigg) \cos \theta \\
&\o_{\phi 24} = - \frac{1}{2r} \chi^{-1} \sin \theta & 
& \o_{\phi 45} = - \Bigg( \half + \frac{2}{{R_4}^2} \chi^{-1} + \frac{2}{r{R_4}^2} \chi^{-2} \Bigg) \cos \theta \\ 
&\o_{\phi 35} = - \Bigg( \half + r \frac{\psi'}{2\psi} + \frac{2}{R_4^2} \chi^{-2} \Bigg) \sin \theta. & 
\label{spin0TN}
\end{align}
Accordingly, for the radial equation we get,
\be
\left( \d_r + \half \o_{r01} {\widetilde{\G}}^{01} - \half {\bar H}^1_{ruv} \G^{uv} \right) \eps = 0.
\ee
Using $\G^{uv} \eps = - 2 \psi \eps$, this equation simplifies to
\be
\left( \d_r + \frac{\psi'}{2 \psi} \right) \e = 0.
\ee

We see that this is exactly the $r$ component of the equation we obtained for the BMPV black hole in section \ref{sec:killing_spinors}. A complete analysis shows that all the other equations also coincide with those obtained for the BMPV black hole---we once again get equations \eqref{KS_x4}, \eqref{KS_th} and \eqref{KS_phi}, independent of the additional factors of $\chi$. Thus, Taub-NUT space has no  effect on the solutions of the Killing spinor equation.

\subsection{Smooth coordinates near the future horizon}
\label{sec:non_singular_coordindate_TN}

Let us start with the non-rotating BMPV black hole in Taub-NUT space.  In the no rotation limit, the metric simplifies to,
\be
ds^2 = \psi^{-1}(dudv + K dv^2 ) + \psi \ ds^2_\rom{TN}, \label{metric_non_rotating_TN}
\ee
where as before,
\begin{equation}
\psi = 1 + \frac{r_{0}}{r} , \qquad K = \psi -1  = \frac{r_0}{r} \, .
\end{equation}
The difference between the above metric and the non-rotating BMPV metric comes only from the replacement of the four-dimensional base from flat space to Taub-NUT space. 

In the $r \to 0$ limit, Taub-NUT space becomes flat space, hence even this difference  disappears in this limit. Thus, as a first guess it is natural to try the same coordinates as for the non-rotating BMPV black hole, i.e., equations \eqref{eq:newcoord1_rev}--\eqref{eq:newcoord3_rev}. When we do this transformation, we find that all terms except for the coefficient of $dV^2$ term are smooth in the $V \to 0$ limit. The $dV^2$ term has a singularity of the form, 
\bea
ds^2 = -\frac{16 r_0^{5/2} W^2}{R_4^2 V} dV^2 + \verb+non-singular terms+.
\eea
This singular term, however, can be easily removed by adjusting the coefficient of  the $ \ln \left(- \frac{\sqrt{r_0}}{V}\right)$ term in  the $ u$ transformation \eqref{eq:newcoord3_rev}. With the transformation, 
\be
 u = U + \frac{1}{2 V W^2} -2 \sqrt{r_0}  \left( 1 + \frac{2 r_0}{R_4^2} \right)  \, \ln \left(- \frac{\sqrt{r_0}}{V}\right), \label{U_new_TN}
\ee
the resulting metric is smooth in the $V \to 0$ limit.  The resulting metric is not particularly illuminating, so we do not present it in full detail. In the $V \to 0$ limit, it takes the form, 
\bea
ds^2 &=& 4 r_0 W^2 dU dV + \frac{32 r_0^2 (8 r_0 + 3 R_4^2) W^4}{R_4^2} dV^2 - \frac{16(4 r_0^{5/2} + 3 r_0^{3/2} R_4^2) W}{R_4^2}  dV  dW \no \\ & & + \ \frac{4 r_0}{W^2} dW^2  + 4 \, r_0 \, d \Omega_3^2 
\eea
This metric is locally $\mathrm{AdS}_3 \times \mathrm{S}^3$. We  verify this by computing the Ricci tensor and then taking the $V \to 0$ limit. For the three-form field strength in new coordinates, we get the same expression as \eqref{three_form_non_rotatingFS}:
\begin{equation}
  F^{(3)} = \frac{r_0}{\lambda}\, \left[ \sin\theta\,
dx^4\wedge d\theta\wedge d\phi + 4WdW\wedge dV\wedge dU\right]\,  .
\end{equation}
In the $V\to 0$ limit, $F^{(3)}$ is well behaved (in fact independent
of $V$). One can also easily check the self-duality property of  $F^{(3)}$.

For the rotating  black hole, metric can also be written in the form:
\be
ds^2 = \psi^{-1}\left(dudv + K dv^2 + \frac{\widetilde J}{4} \left( \frac{1}{r} + \frac{4}{R_4^2}\right)
\, (dx^4+\cos\theta d\phi)
\, dv\right) 
+ \psi ds^2_\rom{TN}. \label{metric_rotating_TN}
\ee
In order to introduce smooth coordinates in the near-horizon region we proceed in steps parallel to section \ref{sec:non_singular_coordindate}. First, we shift $x^4$ coordinate as 
\be
 x^4 = \widetilde x^4 - \frac{\widetilde J}{8r_0^2} v, \label{x4_shift_TN}
\ee
so that the cross term between $d v$ and $(d \widetilde x^4 +\cos\theta d\phi) $ has a zero at $r=0$. The transformed metric takes the form,
\bea
ds^2 &=& \psi^{-1}\left[dudv + \left(K +  \left(\frac{4}{R_4^2} + \frac{1}{r} \right)^{-1} \frac{\widetilde J^2}{64 \, r_0^4}
  \psi^2 - \left(\frac{4}{R_4^2} + \frac{1}{r} \right) \frac{\widetilde J^2}{32 r_0^2} 
\right) dv^2 \right. \no  \\ 
 & & \left. + \left(  \left(\frac{4}{R_4^2} + \frac{1}{r} \right)\frac{\widetilde J}{4}
-  \left(\frac{4}{R_4^2} + \frac{1}{r} \right)^{-1}  \frac{\widetilde J}{4 r_0^2} \psi^2\right)
\, (d\widetilde x^4+\cos\theta d\phi)
\, dv\right]   + \psi \,  \widetilde{ds}^2_\rom{TN}, \no \\
\eea
where now, 
\be
\widetilde{ds}^2_\rom{TN}= \left(\frac{4}{R_4^2} + \frac{1}{r} \right)^{-1} (d \widetilde{x}^4 + \cos \theta \, d\phi)^2 +
\left(\frac{4}{R_4^2} + \frac{1}{r} \right) (dr^2 + r^2 d\theta^2 + r^2 \sin^2\theta \,d\phi^2).
\ee
The shift \eqref{x4_shift_TN} changes the identification under $x^5\equiv x^5+2\pi R_5$.  The new identification takes the form \eqref{new_identification}. Next, we carry out a rescaling,
\begin{align}
 u &= \left(1-\frac{\widetilde J^2}{64r_0^3}\right)^{1/2} \widetilde u, &
 v &=  \left(1-\frac{\widetilde J^2}{64r_0^3}\right)^{-1/2}\widetilde v, 
\end{align}
so that the coefficient of the $d\widetilde v^2$ term in the metric remains unity  as $r\to 0$, as in the non-rotating case. In order to carry out this rescaling we must have, 
\be
\widetilde J^2 <  64r_0^3.
\ee
The rescaling gives,
\bea
ds^2 &=& \psi^{-1}\left[d\widetilde u \ d\widetilde v + \left(K +\left(\frac{4}{R_4^2} + \frac{1}{r} \right)^{-1} \frac{\widetilde J^2}{64 r_0^4} \psi^2
-\left(\frac{4}{R_4^2} + \frac{1}{r} \right) \frac{\widetilde J^2}{32 r_0^2 r}\right)
\left(1- \frac{\widetilde J^2}{64r_0^3}\right)^{-1}d\widetilde v^2\right.
\no \\ 
&& \qquad \left. + \left( \left(\frac{4}{R_4^2} + \frac{1}{r} \right) \frac{\widetilde J}{4}
 - \left(\frac{4}{R_4^2} + \frac{1}{r} \right)^{-1} \frac{\widetilde J}{4 r_0^2}\psi^2\right)\,
 \left(1- \frac{\widetilde J^2}{64r_0^3}\right)^{-1/2}
\, (d\widetilde x^4+\cos\theta \, d\phi)
\, d\widetilde v\right] \no \\
&& + \psi \,  \widetilde{ds}^2_\rom{TN},
\eea

In this metric the coefficient of  $d\widetilde u \ d\widetilde v$ term and the coefficients  of the Taub-NUT coordinates $(r, \widetilde x^4, \theta, \phi)$ are the same as for the non-rotating BMPV black hole. Moreover, in the $r\to 0$ limit,  the coefficients of the $d\widetilde v^2$ and $(d\widetilde x^4+\cos\theta d\phi)
\, d\widetilde v$ terms have the same numerical values as for the non-rotating BMPV black hole. Thus, as a first guess it is natural to try to the same coordinates as for the non-rotating black hole, i.e., equations \eqref{eq:newcoord1_rev_rotating}--\eqref{eq:newcoord3_rev_rotating}. However, as in the previous cases, when we do this transformation, we find that  all terms are smooth in the $V \to 0$ limit except for the coefficient of $dV^2$ term. The $dV^2$ term has a singularity of the form, 
\bea
ds^2 = -  \frac{\widetilde J^2}{8 r_0^{3/2}} \left(1 - 
\frac{6 r_0}{R_4^2}\right) \left(1- \frac{\widetilde J^2}{64r_0^3}\right)^{-1} W^2 \: \frac{dV^2}{V} + \verb+non-singular terms+.
\eea
This singular term can  easily be removed by adjusting the coefficient of  the $ \ln \left(- \frac{\sqrt{r_0}}{V}\right)$ term in   $\widetilde u$ transformation \eqref{eq:newcoord3_rev_rotating}. With the transformation, 
\be
\widetilde u = U + \frac{1}{2 V W^2} - \left( 2 \sqrt{r_0} \left( 1 + \frac{2\, r_0}{R_4^2} \right)
+ \frac{\widetilde J^2}{32 \, r_0^{5/2}} \left(1- \frac{\widetilde J^2}{64 \, r_0^3}\right)^{-1}  \left( 1 - \frac{6 \, r_0}{R_4^2} \right) \right)  \, \ln \left(- \frac{\sqrt{r_0}}{V}\right),
\label{eq:newcoord3_rev_TN_rotating}
\ee
the resulting metric is smooth in the $V \to 0$ limit. The resulting metric is not particularly illuminating, so we do not present those details.

The three-form field strength in the new coordinates is also non-singular. It takes the form, 
\begin{eqnarray}
F^{(3)} &=& \frac{r_0}{\lambda}\, \bigg
[ \sin\theta\, 
d\widetilde x^4\wedge d\theta\wedge d\phi + 4
WdW\wedge dV\wedge  dU   \nonumber \\
&& -  \frac{\widetilde J}{r_0}\, \left(1-\frac{{\widetilde J}^2}
{64 r_0^3}\right)^{-1/2}\,
WdW\wedge dV\wedge
  (d\widetilde x^4 + \cos\theta \,
  d\phi)   \nonumber \\
&&  - \frac{ \widetilde J}{2r_0}\left(1-\frac{{\widetilde J}^2}
{64 r_0^3}\right)^{-1/2}W^2 \, \sin\theta\, 
dV\wedge d\theta\wedge d\phi \bigg]\,  .
\label{eq:three_form_rotatingTN}
\end{eqnarray}
It  can be  confirmed that expression \eqref{eq:three_form_rotatingTN} is self-dual.

\section{Deformations of  the BMPV black hole in Taub-NUT space}
\label{sec:deformations_TN}
In this section, we study deformations of the BMPV black hole in Taub-NUT space. A class of these deformations is generated by Garfinkle-Vachaspati transform. These are studied in section \ref{sec:bosonic_deformations_TN}. Taub-NUT space admits a self-dual harmonic form. A class of deformations is generated by anti-self-dual form fields using this self-dual two-form. These are studied in section \ref{sec:bosonic_deformations_TN_form_fields}.  Finally, a class of fermionic deformations of the type discussed for the BMPV black hole can also be added to the BMPV black hole in Taub-NUT space. These are studied in section \ref{sec:fermionic_deformations_TN}.

\subsection{Bosonic deformations generated by Garfinkle-Vachaspati transform}
\label{sec:bosonic_deformations_TN}

The bosonic deformations generated by the Garfinkle-Vachaspati transform take the form,
\be
ds^2 = \psi^{-1}(r) [du \, dv + (\psi(r) -1 + \widetilde{T}(v, x^4, r, \theta, \phi)) \, dv^2 - 2 \widetilde{\z} \, dv] + \psi(r) \, ds^2_\rom{TN},
\ee
where now the condition is that $\widetilde{T}(v, x^4, r, \theta, \phi)$ is a harmonic function on four-dimensional Taub-NUT space. For an $x^4$ independent function, the condition simply reduces to the function $\widetilde{T}$ being harmonic on three-dimensional transverse space $\mathbb{R}^3$ spanned by $(r, \theta, \phi)$. As in the BMPV case, requiring the deformation to be regular at the origin and at infinity and dropping terms that can be removed by coordinate transformations, we can choose
\be
\widetilde{T}(v, \vec y) = g_i(v) \, y^i, \qquad \qquad \int_0^{2\pi R_5} g_i(v) \, dv = 0,
\ee
where $y^i$ are cartesian coordinates on $\mathbb{R}^3$ and $g_i(v)$ are three arbitrary functions of $v$  apart from the restriction that their integral over $dv$ is zero (this eliminates the constant terms in the Fourier series of $g_i(v)$). The metric is not manifestly asymptotically flat, but using a standard set of coordinate transformations, it can be brought into a manifestly asymptotically flat form. 
 
In the $\rho, \tau,v$ near-horizon coordinates, these bosonic deformations scale as $\beta^2$ in the $\beta \rightarrow 0$ limit. 
To characterise these deformations into smooth hair modes or not, we now turn to their smoothness analysis. The smoothness analysis below generalises the corresponding discussion of \cite{Jatkar:2009yd} to rotating black holes.
  
The deformation adds the following extra term to the metric,
\be
\delta (ds^2) =  \psi^{-1} \, \widetilde{T}(v, \vec y) \: dv^2 = \psi^{-1} \, g_i(v) \, y^i \: dv^2 =  r \, \psi^{-1} \, g_i(v) \, n^i \: dv^2 \label{deform_TN} ,
\ee
where $n^i = y^i/|y|$ in the three-dimensional unit vector. The SO(3) unit vector $n^i$ only depends on the angular coordinates. Let us first set $\widetilde{J} =0$. For the non-rotating BMPV black hole in Taub-NUT space, a non-singular set of coordinates  are \eqref{eq:newcoord1_rev}, \eqref{eq:newcoord2_rev}, and \eqref{U_new_TN}. In these coordinates, the extra term takes the form, 
\be
\delta (ds^2) = 16 r_0^3 \, n^i \, g_i(v) \, W^4 (1+ 4 \sqrt{r_0} W^2 V)^{-1} \, dV^2. \label{singular_TN}
\ee
As $V \to 0$, $v$ coordinate changes rapidly from a finite value to infinity. As a result, $g_i(v)$, although finite at the horizon, oscillate  rapidly as $V \to 0$. We want to get rid of these rapid oscillations. We can ensure that $\delta G_{VV}$ vanishes by a shift in the $U$ coordinate
\be
U = \widetilde U - H(V, W,  \theta, \phi),
\ee
with
\be
H(V, W,  \theta, \phi) = 4 r_0^2 W^2 \int_0^V  (1+ 4 \sqrt{r_0} W^2 V')^{-1} n^i g_i(v(V'))  dV'.
\ee
The shift generates additional terms in the metric,
\bea
& & - 4 r_0 W^2 \partial_W  H(V, W, \theta, \phi)  dV dW - 4 r_0 W^2  \partial_{\theta_i}  H(V, W, \theta, \phi)  d \theta^i  dV
\eea
where $\theta^i$ collectively denotes $\theta, \phi$. These additional terms all vanish in the $V \to 0$ limit. The resulting metric is smooth at $V=0$; however,  $V$ derivatives of the function $H$ are not. Specifically, although $\partial_V H$ is finite at $V=0$, $\partial_V^2 H$ diverges at $V=0$. These divergences, however, do not appear in the Riemann tensor components. It can be seen by an explicit calculation. It can also be argued using weight considerations. The divergent terms necessarily have weight 3: $\partial_V^2 G_{WV}, \partial_V^2 G_{V \theta_i} $.  Whereas covariant Riemann tensor components are of at most weight 2. Hence, such divergent terms do not appear in the Riemann tensor components.

  For the rotating BMPV black hole in Taub-NUT, a similar analysis applies. A non-singular set of coordinates  are  \eqref{eq:newcoord1_rev_rotating}, \eqref{eq:newcoord2_rev_rotating}, \eqref{x4_shift_TN}, and \eqref{eq:newcoord3_rev_TN_rotating}. In these coordinates, the extra term \eqref{deform_TN} takes the form, 
\be
\delta (ds^2) = \frac{1024  }{64 r_0^3-\widetilde{J}^2}   r_0^{6} W^4 (1+ 4 \sqrt{r_0} W^2 V)^{-1}  n^i g_i(v) \: dV^2. \label{singular_TN_rotating}
\ee
As $V \to 0$, $v$ coordinate changes rapidly from a finite value to infinity. Once again, we can ensure that $\delta G_{VV}$ vanishes by a  shift in the $U$ coordinate
\be
U = \widetilde U - \widetilde{H}(V, W,  \theta, \phi),
\ee
with
\be
 \widetilde{H}(V, W, \theta, \phi) =  \frac{256 r_0^{5}}{64 r_0^3- \widetilde{J}^2} W^2  \int_0^V (1 + 4 \sqrt{r_0} V' W^2)^{-1} n^ig_i(v(V')) dV' .
\ee
The shift generates 
\bea
& & - 4 r_0 W^2 \partial_W  \widetilde{H}(V, W, \theta, \phi)  dV dW - 4 r_0 W^2  \partial_{\theta_i}  \widetilde{H}(V, W, \theta, \phi) dV d \theta^i 
\eea
where $\theta^i$ collectively denotes $\theta, \phi$.  These additional terms once again all vanish in the $V \to 0$ limit. The resulting metric is smooth at $V=0$; however, $V$ derivatives of the function $\widetilde{H}$ are not. Specifically,  $\partial_V^2 \widetilde{H}$ diverges at $V=0$. These divergences, however, do not appear in the Riemann tensor components.  
 
Thus, we see that the three functions $g_i(v)$ generate smooth deformations of the BMPV black hole in Taub-NUT. The modes are supported entirely outside the horizon. They are genuine hair modes.

\subsection{Two-form field deformations}
\label{sec:bosonic_deformations_TN_form_fields}
Taub-NUT space has a self-dual harmonic two-form (with convention $\epsilon_{x^4 r\theta \phi} = +1$),
\be
\omega_\rom{TN} = - \frac{r}{4 r + R_4^2} \sin \theta \: d\theta \wedge d\phi + \frac{R_4^2}{(4r + R_4^2)^2} dr \wedge (dx^4 + \cos \theta \, d \phi).
\ee
Using this two form, a six-dimensional anti-self-dual three-form can be constructed by simply 
\be
h^s(v)\, dv \wedge \omega_\rom{TN}
\ee
where $ h^s(v)$ is an arbitrary function of $v$ for $1 \le s \le n_t$. Since in the $(2,0)$ theory under consideration, there are  $n_t$ tensor multiplets, there are $n_t$ such deformations.  Such a term can be taken to be the source for deformation in the tensor-multiplet sector, with
\be
\delta H^s = h^s(v) \, dv \wedge \omega_\rom{TN}.
\ee

Taking the deformation of this form, bosonic equation \eqref{eeom2} continues to be satisfied with the self-dual three-form supporting Taub-NUT black hole \eqref{F3_TN}. Equation \eqref{eeom1}
is satisfied with metric deformed by 
\be
ds^2 = \psi^{-1}(r) [du dv + (\psi(r) -1 + \widetilde{S}(v, r)) dv^2 - 2 \widetilde{\z} dv] + \psi(r) \: ds^2_\rom{TN},
\ee
with function $ \widetilde{S}(v, r)$ satisfying
\be
\nabla^2_\rom{TN} \widetilde{S}(v, r) = \frac{R_4^2}{{4 r+R_4^2}} \left(r \,  \partial_r^2 \widetilde{S} (v,r) + 2 \, \partial_r \widetilde{S}(v,r)\right) = -    \frac{8R_4^4}{(4 r+R_4^2)^4} \left(\sum_{s=1}^{n_t}  h^s(v) h^s(v)\right),
\ee
where $\nabla^2_\rom{TN} $ is the Taub-NUT Laplacian. A solution for the function $\widetilde{S}$ can be taken to be, 
\be
\widetilde{S} (v,r) = - \frac{4r}{R_4^2(4 r+R_4^2)} \left(\sum_{s=1}^{n_t}  h^s(v) h^s(v)\right).
\ee

The function $\widetilde{S} (v,r)$ does not vanish at infinity, so once again the deformed metric does not look manifestly asymptotically flat. However, this can be readily fixed by shifting $u$ coordinate as,
\be
u \to u +  \frac{1}{R_4^2} \int^v_0 \left( \sum_{s=1}^{n_t} h^s(v') h^s(v')\right) dv'.
\ee

The anti-self-dual form fields $ \delta H^s $ in the  near-horizon limit \eqref{NHcoord} scale as $\beta$ and the metric deformation term $\psi^{-1}(r) \widetilde S(v,r) $ scales as $\beta^2$.  Since these terms vanish in the near-horizon $ \beta \to 0$ limit, they are potential hair modes.

How the supersymmetry properties are affected due to these deformations? We note that the Killing spinor equations with the anti-self-dual form fields turned on are \eqref{del_grav} and \eqref{del_spin1/2}. An anti-self-dual form field deformation enters directly in  equation \eqref{del_spin1/2} and in equation \eqref{del_grav} through the weight 2 term added to the metric. As we also argued in section \ref{sec:bosonic_deformations}, a weight 2 term in the metric does not alter the Killing spinor analysis for equation  \eqref{del_grav}.  The anti-self-dual form field deformation being weight 1, necessarily come as $\G^{vij}H^s_{vij} \eps$ in equation \eqref{del_spin1/2}. All such terms vanish due to  the projection condition $\Gamma^v \epsilon =0$  for $\eps$ used in section \ref{sec:killing_spinors_TN}. Thus, the anti-self-dual form field deformations preserve  the supersymmetry properties of the original background.

To check whether the deformation is smooth or not at the future horizon, we follow the same procedure as before. We write the deformed solution in  smooth coordinates of the background spacetime and check whether the deformation is well behaved or not. If not, then we perform coordinate transformations to make the deformation well behaved and check for the Riemann tensor and  matter field strength components. If the metric, matter field strength, and Riemann tensor components all turn out to be smooth, we declare that  the deformation is smooth. 

The deformed anti-self-dual three-form field in the new coordinates near $V=0$ behaves as
\be
\delta H^s \simeq \frac{32 r_0^{7/2}h^s(v)}{R_4^2 (64r_0^3 - \widetilde{J}^2)^{1/2}} d V \wedge \left( - W^2 \,   \sin \theta \, d \theta \wedge d \phi + 2 W  \,  dW  \,  \wedge  \,  (dx^4 + \cos \theta \, d \phi) \right). \label{three_form_TN_BH}
\ee
This tensor is clearly smooth near the horizon $V=0$. The functions $h^s(v)$ depend on the $v$ coordinate that changes  from a finite value to infinity at the horizon. As a result, $h^s(v)$, although finite at the horizon, oscillate rapidly as $V \to 0$. Since we do not need to take further derivatives on this function, this is not a concern.

The deformation term in the metric in the new coordinates becomes,
\bea
\delta \left(ds^2\right) &=&  \psi^{-1}(r) \widetilde{S}(v, r)) dv^2 \ = \ -  \frac{4 r \, \psi^{-1}(r)}{R_4^2(4 r+R_4^2)} \left(\sum_{s=1}^{n_t} \,  h^s(v) h^s(v)\right)  dv^2 \label{TN_metric_def} \\
&=& - 8 \left(\sum_{s=1}^{n_t}  h^s(v) h^s(v)\right)    \frac{512 \, r_0^6 \,  W^4 (64 r_0^3- \widetilde{J}^2)^{-1}}{R_4^2(R_4^2 (1 + 4 \sqrt{r_0} V W^2)- 16 r_0^{3/2} V W^2)} dV^2. \quad
\eea
As $V \to 0$, $v$ coordinate changes  from a finite value to infinity. As a result, $h^s(v)$ oscillate  rapidly as $V \to 0$. We can ensure that $\delta G_{VV}$ vanishes by a shift in the $U$ coordinate
\be
U = \widetilde U - F(V, W), \label{change_of_coordinate}
\ee
with
\be
 F(V, W) =  -   \frac{1024 \, W^2 \, r_0^{5}}{R_4^2(64 r_0^3- \widetilde{J}^2)}  \int_0^V \frac{ \sum_{s=1}^{n_t} h^s(v(V')) h^s(v(V'))}{(R_4^2 (1 + 4 \sqrt{r_0} V' W^2)- 16 r_0^{3/2} V' W^2)} dV' .
\ee
The shift generates an additional term in the metric, 
\bea
& & - 4 r_0 W^2 \partial_W  F(V, W) dW dV.
\eea This additional term vanishes in the $V \to 0$ limit. The resulting metric is thus smooth near $V=0$; however, the $V$ derivatives of the function $F$ are not. Specifically,  $\partial_V^2 F$ diverges at $V=0$. These divergences, however, do not appear in the Riemann tensor components: the divergent terms are of weight 3. The change of coordinate  \eqref{change_of_coordinate} does not affect the three-form \eqref{three_form_TN_BH}. 

Thus, for each tensor multiplet, there is a smooth deformation parameterised by an arbitrary function $h^s(v)$. 

\subsection{Fermionic deformations}
\label{sec:fermionic_deformations_TN}
The construction of fermionic deformations closely follows the logic of section \ref{sec:fermionic_deformations}. We again use the ansatz and gauge condition \eqref{ansatz+gauge} and by similar arguments utilizing the concept of weight, arrive at the weight 1 equation \eqref{grav-wt1} to be solved in the Taub-NUT background. Since $\Psi_v$ is already weight 1, we only need weight 0 contributions from the other fields. The appropriate weight 0 truncated vielbeins are given in equations \eqref{frame0TN01}--\eqref{frame0TN45}
and the non-zero spin connection coefficients  are given in \eqref{spin0TNr} - \eqref{spin0TN}.
Substituting these into the  gravitino equation \eqref{grav-wt1}, with the function $\chi(r)$ defined in \eqref{TN function}, we get,
\bea
& (\psi \chi)^{-1/2} \widetilde{\Gamma}^3&\Bigg( \partial_r +\frac{\psi'}{\psi} + \frac{1}{r} - \half \frac{\psi'}{\psi} \widehat{\Gamma}^1 \Bigg) \Psi_v + \frac{(\psi \chi)^{-1/2}}{{r \sin\theta}} \widetilde{\Gamma}^5  \partial_{\phi} \Psi_v \no \\ 
&+ \psi^{-1/2}\chi^{1/2}&\Bigg(\widetilde{\Gamma}^2 - \frac{\cot\theta}{r \chi} \widetilde{\Gamma}^5 \Bigg) \partial_{x^4} \Psi_v + \frac{(\psi \chi)^{-1/2}}{r} \widetilde{\Gamma}^4 \Bigg( \partial_{\theta} + \half \cot\theta \Bigg) \Psi_v = 0.
\eea
As was the case in flat space, we take fermionic modes to have no momentum  along the $x^4$ direction, i.e., $\partial_{x^4} \Psi_v =0.$  Then,  the common factor of $\chi^{-1/2}$ means that the gravitino equation reduces to the radial and angular equations that result from \eqref{grav eqn}. Thus, the Taub-NUT space does not have any effect on the fermionic deformations obtained earlier. 

In the near-horizon limit, these deformations  scale as $\beta^{3/2}$ just like their flat space counterparts.

In section \ref{sec:bosonic_deformations_TN_form_fields}, we switched on anti-self-dual form fields that could enter the second gravitino equation \eqref{H-gravitino}. This is a scalar equation, i.e., weight 0, whereas all the form field deformations $H_{s MNP}$ that were turned on are of weight 1. This means that all the non-zero components of $H^{s MNP}$  have one $u$ index and no $v$ index. We must have a $v$ index in order to contract with $\Psi_v$. Since this is not possible, it follows that the second fermionic equation is identically satisfied. Previously it was noted that the weight 2 terms in the metric do not affect the fermionic deformation modes. Hence we do not need to worry about the effect of metric deformations \eqref{deform_TN} or \eqref{TN_metric_def} on the fermion modes. 

Like their flat space counterparts, these fermionic deformations preserve the supersymmetry of the original background.

The smoothness analysis also proceeds as in the case of flat space. We begin by looking at the vielbeins in the smooth coordinates and find that vielbeins \eqref{frame2FS_TN}--\eqref{frame2FS5_TN} are singular. We analyse the non-rotating and  rotating cases separately. For non-rotating BMPV black hole in Taub-NUT,
\bea
e^+ &:=& e^0 + e^1 = - \sqrt{r_0} \frac{dV}{V},\\
e^- &:=& e^1 - e^0 =  \sqrt{r_0} \frac{dV}{V}  - 12 r_0 W^2 \left(1+ \frac{4 r_0}{3 R_4^2} \right) dV -4 \sqrt{r_0} V W^2 dU + \frac{4 \sqrt{r_0}}{W} dW, \quad  
\eea
\bea
e^2 &=& \sqrt{r_0} \left(1 + 4 \sqrt{r_0} \left(1- \frac{4 r_0}{R_4^2} \right) V W^2\right)^{-1/2} \left( d x^4 + \cos \theta \, d \phi  \right), \\
e^3 &=&  \frac{2 \sqrt{r_0}}{(1+ 4 \sqrt{r_0} V W^2)^{2}} \left(1 + 4 \sqrt{r_0} \left(1- \frac{4 r_0}{R_4^2} \right) V W^2 \right)^{1/2} \left( \frac{dW}{W} + \frac{dV}{2V}\right) ,
\eea
\bea
e^4 &=& \frac{\sqrt{r_0}}{(1+ 4 \sqrt{r_0} V W^2)}\left(1 + 4 \sqrt{r_0} \left(1- \frac{4 r_0}{R_4^2} \right) V W^2 \right)^{1/2} d\theta,\\
e^5 &=& \frac{\sqrt{r_0}}{(1+ 4 \sqrt{r_0} V W^2)} \left(1 + 4 \sqrt{r_0} \left(1- \frac{4 r_0}{R_4^2} \right) V W^2 \right)^{1/2}\sin \theta \, d \phi.
\eea
Note that $e^+, e^-, e^3$ are singular at $V=0$. The metric in non-singular coordinates can be expressed as
\be
ds^2 = e^+ e^- + (e^2)^2 +(e^3)^2 + (e^4)^2 + (e^5)^2  .
\ee

A non-singular set of vielbeins can be obtained by a sequence of Lorentz transformations: first, 
\begin{align}
& \hat e^+ = -2 V e^+, &
& \hat e^- = - \frac{1}{2V} e^-, 
& \hat e^3 = e^3,
\end{align}
and then, 
\begin{align}
& \tilde  e^+ = \hat e^+,&
& \tilde e^- = \hat e^- - 2 \beta \hat  e^3 - \beta^2 \hat e^+,
& \tilde e^3 = \hat  e^3 + \beta \hat e^+,
\end{align}
with
\be
\beta=-\frac{1}{2 V (1+ 4 \sqrt{r_0} V W^2)^{2}}\left(1 + 4 \sqrt{r_0} \left(1- \frac{4 r_0}{R_4^2} \right) V W^2 \right)^{1/2}. \label{beta_TN}
\ee

For the rotating black hole in Taub-NUT, the corresponding expressions are much more cumbersome. We refrain from presenting those details. It suffices to say that  in the new coordinates $e^+, e^-, e^2, e^3$ are singular at $V=0$. A non-singular set of vielbeins can be obtained by a sequence of Lorentz transformations, as before: first, 
\begin{align}
& \hat e^+ = \alpha e^+, &
& \hat e^- = \frac{1}{\alpha} e^-, 
& \hat e^2 = e^2, &
& \hat e^3 = e^3,
\end{align}
then, 
\begin{align}
& \check e^+ = \hat e^+,&
& \check e^- = \hat e^- - 2 \beta \hat  e^3 - \beta^2 \hat e^+,
& \check e^2 = \hat e^2, &
& \check e^3 = \hat  e^3 + \beta \hat e^+,
\end{align}
and finally, 
\begin{align}
& \tilde e^+= \check e^+ ,&
& \tilde e^- = \check e^- - 2 \gamma \check e^2 - \gamma^2 \check e^+,
& \tilde e^2 = \check e^2 + \gamma \check e^+ ,&
& \tilde e^3 = \check e^3,
\end{align}
where
\bea
\alpha&=&- \frac{(64 r_0^3-\widetilde{J}^2)^{1/2}}{4 r_0^{3/2}} V, \\
\beta&=&-\frac{1}{2 V (1+ 4 \sqrt{r_0} V W^2)^{2}}\left(1 + 4 \sqrt{r_0} \left(1- \frac{4 r_0}{R_4^2} \right) V W^2 \right)^{1/2}, \\
\gamma &=& -\frac{\widetilde{J}}{2 V  (64 r_0^3-\widetilde{J}^2)^{1/2}}\left(1 + 4 \sqrt{r_0} \left(1- \frac{4 r_0}{R_4^2} \right) V W^2 \right)^{-1/2}.
\eea
 
  These local Lorentz transformations act on the gravitino field. As shown in appendix \ref{app:Lorentz} the combined action is simply,
 \be
 \widetilde{\Psi}_V = \widehat{\Psi}_V = \frac{1}{\sqrt{\alpha}} \Psi_V = \frac{128 r_0^{7/2}}{ (64 r_0^3-\widetilde{J}^2)^{3/4}}  \, W^3 \, \eta(v(V), \theta, \phi).
 \ee
The field is smooth in the $V \to 0 $ limit. This field does not vanish at the horizon. However, we can make it vanish by doing a local supersymmetry transformation with a parameter proportional to 
 \be
  W^3 \,  \int^V_0  \eta(v(V'), \theta, \phi) dV'.
 \ee
 
   \numberwithin{equation}{subsection}
\section{Hair removed 4d and 5d partition functions}
\label{sec:microscopics}
Having obtained the  hair modes as solutions to non-linear supergravity equations for both 4d and 5d black holes, we now turn to the discussion of hair removed  partition functions. The hair removed 4d and 5d partition functions themselves are interesting quantities, as they are expected to be obtainable on the gravity side from the quantum  entropy function formalism~\cite{Sen:2008vm, Mandal:2010cj, Sen:2014aja}. In section \ref{4d_5d_counting_formulas}, we review the microscopic considerations relevant for our discussion.  
In section \ref{hair_partition_functions}, we identify twisted sectors hair modes in ten-dimensional supergravity description and compute the hair removed 4d and 5d partition functions. In section \ref{matching}
we show the hair removed 4d and 5d partition functions perfectly match.

\subsection{4d and 5d counting formulas}
\label{4d_5d_counting_formulas}
We consider type IIB string theory compactified on $\cM \times \widetilde{\mathrm{S}}^1 \times \mathrm{S}^1$ where $\cM$ is either K3 or T$^4$ and mod out this theory by a $\mathbb{Z}_{N}$ symmetry generated by $1/N$ shift along the S$^1$ and an order $N$ transformation $\widetilde g$ on $\cM$. The orbifolding acts in such a way that the final theory has 16 real supercharges, i.e., ${\cal N}=4$ supersymmetry in four-dimensions, equivalently (2,0) supersymmetry in six-dimensions.  These models \cite{Chaudhuri:1995fk, Chaudhuri:1995bf, Schwarz:1995bj, Chaudhuri:1995dj} are widely studied in the context of precision counting of black hole microstates~\cite{hep-th/9607026, hep-th/0503217, hep-th/0505094, hep-th/0510147, hep-th/0602254, hep-th/0605210, hep-th/0609109,  Sen:2007vb, Dabholkar:2007vk,  Cheng:2007ch, Dabholkar:2008zy, Banerjee:2009uk,
Jatkar:2009yd, 0907.1410, 1109.3706}; and are reviewed in \cite{0708.1270}. Most of our comments below are for $\cM =$ K3, some comments are for T$^4$.

We follow the notation and conventions of \cite{0708.1270}. We take the radius of S$^1$ to be $N$ and the radius of $ \widetilde{\mathrm{S}}^1$ to be 1. In this convention the orbifolded circle $\mathrm{S}^1/\mathbb{Z}_{N}$ has radius 1. The action of the orbifolding group involves translations along the $ \mathrm{S}^1$ by $2 \pi$ and under this translation various fields get transformed by a $\widetilde g$ action. Momentum along the circle S$^1$ is quantised in units of $1/N$. Following \cite{hep-th/0505094, 0708.1270}, we consider a single D5 brane wrapped on $\cM \times \mathrm{S}^1$, $Q_1$ D1-branes wrapped on S$^1$, a single KK monopole with negative charge associated with the circle $\widetilde{\mathrm{S}}^1$ and momentum $-n/N$ along the S$^1$ and momentum $J$ along $\widetilde{\mathrm{S}}^1$. Since the D5 brane wraps the four-dimensional manifold $\cM$,  it also carries a negative D1 charge given  by the Euler character $\chi(\cM)$ of $\cM$ divided by 24 \cite{Bershadsky:1995qy}. The net D1 charge is therefore, $Q_1 - \beta$, where 
\be
\beta = \frac{1}{24}\chi(\cM).
\ee
For such a set-up T-duality invariant charge bilinears are
\begin{align}
Q^2 &= 2n/N, & P^2 &= 2 (Q_1 - \beta), & Q \cdot P = J.
\end{align}

 Let us denote by $d_\rom{4d}(n, Q_1, J)$ the helicity trace index for the four-dimensional black hole carrying charges $(n, Q_1, J)$.  The four-dimensional partition function is defined as
\be
Z_\rom{4d}(\widetilde \rho,\widetilde \sigma, \widetilde v) = \sum_{n, Q_1, J} 
(-1)^{J+1} d_\rom{4d}(n, Q_1, J) \exp[2 \pi i \{(Q_1-\beta)/N \widetilde \sigma + n \widetilde \rho + J \widetilde v \} ].
\ee
In the region of the moduli space where the type IIB string coupling is small, the result for $d_\rom{4d}(n, Q_1, J)$ for the models we consider is 
\be\label{4d}
d_\rom{4d}(n, Q_1, J)= (-1)^{Q\cdot P+1}\,
{1\over N}\, \int _\cC d\widetilde\rho \, 
d\widetilde\sigma \,
d\widetilde v \, e^{-\pi i ( N\widetilde \rho Q^2
+ \widetilde \sigma P^2/N +2\widetilde v Q\cdot P)}\, \, Z_\rom{4d} (\widetilde \rho,\widetilde \sigma, \widetilde v),
\ee
with
\be
Z_\rom{4d} (\widetilde \rho,\widetilde \sigma, \widetilde v) = {1
\over \widetilde\Phi(\widetilde \rho,\widetilde \sigma, \widetilde v)},
\ee
where details on the contour $\cC$ and an explicit expression for   $\widetilde\Phi(\widetilde \rho,\widetilde \sigma, \widetilde v)$ are given in equations (5.1.4) and (5.1.5) of \cite{0708.1270}.

The index  $d_\rom{4d}(n, Q_1, J)$ of 1/4-BPS states in the four dimensional
theory was obtained \cite{hep-th/9607026,hep-th/0510147,hep-th/0605210} by placing the  five dimensional  D1-D5 
system in the background of an $\widetilde{\mathrm{S}}^1$ Kaluza-Klein
monopole \cite{hep-th/0503217}. To obtain the index for the five-dimensional system,  we simply
need to \cite{1109.3706}\footnote{We thank Nabamita Banerjee for very helpful discussion on these issues.},
\begin{itemize}
\item remove from the index of the four dimensional system the  contribution of the Kaluza-Klein monopole,
\item remove the contribution of the supersymmetric quantum mechanics that
binds the D1-D5 system to the Kaluza-Klein monopole,
\item and 
multiply with the contribution of fermion
zero modes present in the five dimensional 
system.
\end{itemize}

This procedure gives us a 5d partition function, $\widetilde{Z}_\rom{5d}(\widetilde \rho,\widetilde \sigma, \widetilde v)$. However, this is not the final answer. There is an additional subtlety~\cite{0705.1847,0807.0237,0807.2246}. The 5d electric charges measured at infinity differ from the charges measured at the horizon. This difference arises due to the inclusion of higher derivative Chern-Simons coupling in the 5d action. In our context, this effect amounts to  producing a shift of $\frac{c_2}{24}$ units in the left momentum charge along the S$^1$, where $c_{2}$ is the second Chern class for the compactification manifold $\cM \times \widetilde{\mathrm{S}}^1 \times \mathrm{S}^1$. In the unorbifolded theory with $\cM =$ K3, $c_{2}$ is 24 and hence this effect produces a shift of the S$^1$ momentum by one unit. It implies that if the 5d black hole carries $-n/N$ units of  momentum in the orbifolded theory then the corresponding 4d black hole carries $-(n-1)/N$ units of momentum. 
The 5d partition function is hence defined as
\be
Z_\rom{5d}(\widetilde \rho,\widetilde \sigma, \widetilde v) = \sum_{n, Q_1, J} 
(-1)^{J+1} d_\rom{5d}(n, Q_1, J) \exp[2 \pi i \{(Q_1-\beta)/N \widetilde \sigma + (n-1) \widetilde \rho + J \widetilde v \} ],
\ee
 where $d_\rom{5d}(n, Q_1, J)$ is the modified helicity trace index 
for the five-dimensional black hole carrying charges $(n,Q_1,J)$. For the definition of modified helicity trace index see \cite{Banerjee:2009uk}. 
Note that the coefficient of $\widetilde \rho$ is $(n-1)$ in the above equation. 
We conclude that 
\be
Z_\rom{5d}(\widetilde \rho,\widetilde \sigma, \widetilde v) = \widetilde{Z}_\rom{5d}(\widetilde \rho,\widetilde \sigma, \widetilde v) \, e^{-2 \pi i \widetilde \rho}. \label{shifted_partition}
\ee

To compute $\widetilde{Z}_\rom{5d}(\widetilde \rho,\widetilde \sigma, \widetilde v)$, we proceed as discussed above.
The partition function associated with the
supersymmetric quantum mechanics that describes the D1-D5 center
of mass motion in the KK monopole background is \cite{0708.1270},
\be
-(e^{\pi i \widetilde v}-
e^{-\pi i \widetilde v})^{-2}.
\ee
The additional zero modes present in the five dimensional system contribute~\cite{Banerjee:2009uk,1009.3226}
\be
-(e^{\pi i \widetilde v}-
e^{-\pi i \widetilde v})^{2}.
\ee
As a result, 
\be
\widetilde{Z}_\rom{5d}(\widetilde \rho,\widetilde \sigma, \widetilde v) = 
(e^{\pi i \widetilde v}-
e^{-\pi i \widetilde v})^4 \frac{Z_\rom{4d} (\widetilde \rho,\widetilde \sigma, \widetilde v)}{Z_\rom{KK}(\widetilde \rho)}
=
(e^{\pi i \widetilde v}-
e^{-\pi i \widetilde v})^4 {f(\widetilde\rho)\over \widetilde\Phi(\widetilde\rho,\widetilde\sigma,\widetilde v)} \, ,
\ee
where $1/f(\widetilde\rho)$ is the partition function associated
with a single Kaluza-Klein monopole. An expression for $f(\widetilde\rho)$ is~\cite{hep-th/0609109, 0708.1270}\footnote{This is equation (5.2.13) of the review \cite{0708.1270}. To simplify the presentation (see also e.g.~\cite{1109.3706}) we have removed the factor of 16 associated with the degeneracy coming from the fermion zero modes of the KK monopole in type IIB theory. This is not a mistake. This factor is properly taken into account in the full partition functions.}, 
\be\label{KKpartitionT4}
f(\widetilde\rho) 
=e^{-2\pi i \widetilde \alpha\widetilde\rho}\, \prod_{l=1}^\infty (1 -  
e^{2\pi i l\widetilde\rho})^{-n_l},
\ee
where
\be\label{ecvalue}
C = -\widetilde \alpha/N\, , \qquad  C = -{1\over 24} \sum_{l=0}^{N-1} \, n_l + {1\over 4}\, 
 \sum_{l=0}^{N-1} \, n_l \, {l\over N} \, \left( 1 -{l\over N}\right),
\ee
and $n_l$ given as
\be\label{n_l}
n_l =  
\sum_{s=0}^{N-1}
e^{-2\pi i l s/N}
\, Q_{0,s},
\ee
with
$Q_{0,s}$ being the number of left handed 
bosons minus fermions weighted by $\widetilde g^s$, on the world volume of KK monopole.
 The number  $C$ represents the momentum quantum number in units of $1/N$ of the vacuum of the Kaluza-Klein monopole when all oscillators are in their ground states. 
 
 For the K3$/\mathbb{Z}_N$ models with $N = 1,2,3,5,7$ (non-composite numbers) $Q_{0,s}$ can be read from~\cite{hep-th/0605210} and for $N=4, 6, 8$ (composite numbers) they can be read from \cite{0907.1410}. These numbers are summarised in table \ref{table1}.   Substituting these values give the functions $f(\widetilde\rho)$ for different $N$ in simplified form as products of scaled Dedekind $\eta-$functions.  These products of   $\eta-$function are most  conveniently described in terms of the  associated cycle shape for orbifolds of K3, which we now briefly discuss following \cite{0907.1410}.

 \begin{table}[t!]
\begin{center}
\begin{tabular}{|c||c|c|c|c|c|c|c|c|c|c|c|} \hline
$N$ & $Q_{0,0}$ & $Q_{0,1}$ & $Q_{0,2}$ & $Q_{0,3}$ & $Q_{0,4}$ & $Q_{0,5}$ & $Q_{0,6}$ & $Q_{0,7}$ & $\widetilde{\alpha}$ & $k+2$ & $n_t + 3$\\ \hline \hline
1&24&  &  & & &  & & & 1 &12 & 24  \\ \hline
2&24&8&  & & &  & & & 1 & 8 & 16 \\ \hline
3&24&6&6&  &  &  & & & 1 & 6 & 12 \\ \hline
4&24 &4&8  &4  &  & &  &  & 1 & 5&10 \\ \hline
5&24  &4&4&4&4&  & & & 1& 4 & 8\\ \hline
6&24 &2 &6  &8  &6  &2 &  &  & 1 & 4&8 \\ \hline
7&24&3&3&3&3&3&3 & & 1 & 3 & 6\\ \hline
8&24 &2 & 4 &2  & 8  &2 &  4 &2  & 1 &3 & 6 \\ \hline
\end{tabular}
\caption{ \sf Useful data on $\mathbb{Z}_N$ orbifolds of $ \text{K3} \times \mathrm{S}^1 \times \widetilde{\mathrm{S}}^1$ for calculating 5d partition function from 4d partition function. Note from the last two columns that $2(k+2) = n_t + 3$.}
\label{table1}
\end{center}
\end{table}

  One associates a cycle shape   to an orbifold of K3. A cycle shape is of the form
\be
\rho=1^{a_1}2^{a_2}\cdots N^{a_N},
\ee
with
\be
\sum_j j\, a_j = 24.
\ee 
They are written in table \ref{table_2}. Given a cycle shape, the function $f(\widetilde \rho)$ is given by the eta product as follows:
\begin{equation}
f(\widetilde\rho) = \prod_{j=1}^N \eta(j \, \widetilde\rho)^{a_j}\ .
\end{equation}
This is a modular form of a level $N$ subgroup of PSL$(2,\mathbb{Z})$ of weight \be
w=\frac12 \sum_j a_j:=k+2.
\ee
Thus, for $N = 1,2,3,5,7$,
\be \label{KKpartitionK3}
f(\widetilde\rho) = \eta(\widetilde\rho)^{k+2} \eta(N\widetilde\rho)^{k+2},
\ee
with 
\be
k + 2 = \frac{24}{N+1}.
\ee
For,
\begin{align} 
\label{KKpartitionK3}
N &=4 & k &= 5 &   f(\widetilde\rho) &= \eta(\widetilde\rho)^{4} \eta(2\widetilde\rho)^{2} \eta(4\widetilde\rho)^{4}, \\
N &=6& k &= 4 &   f(\widetilde\rho) &= \eta(\widetilde\rho)^{2} \eta(2\widetilde\rho)^{2} \eta(3\widetilde\rho)^{2} \eta(6\widetilde\rho)^{2}, \\
N &=8 & k &= 3&  f(\widetilde\rho) &= \eta(\widetilde\rho)^{2} \eta(2\widetilde\rho) \eta(4\widetilde\rho)  \eta(8\widetilde\rho)^{2}.
\end{align} 
The number of tensor multiplets $n_t$ associated with these compactifications is given as 
\be
n_t + 3 =\sum_j a_j  = 2 (k+2). \label{number_nt}
\ee

We note that for $N=5$ and $N=6$  the number of tensor multiplets in six-dimensional supergravity description are the same, $n_t = 5$. In contrast, the functions  $f(\widetilde \rho)$ are different. Thus, it is clear that just knowing the number of tensor multiplets is not enough to fix the function  $f(\widetilde \rho)$ uniquely. The same is true for $N=7$ and $N=8$ where $n_t= 3$.

 For $\mathbb{Z}_N$ orbifolds of $\text{T}^4 \times \mathrm{S}^1 \times \widetilde{\mathrm{S}}^1$ the shift in the partition function \eqref{shifted_partition} is not there. Thus,
\be
Z_\rom{5d}(\widetilde \rho,\widetilde \sigma, \widetilde v) = \widetilde{Z}_\rom{5d}(\widetilde \rho,\widetilde \sigma, \widetilde v). \label{no_shift}
\ee
 The Frame shapes for $\text{T}^4/\mathbb{Z}_N$ models for $N=2,3,4$ are shown in table \ref{table3} in the D1-D5 frame.\footnote{This table is incorrect in the published version of the paper, JHEP \textbf{02}, 125 (2021).}

\begin{table}[htb]
\centering
\begin{center}
\begin{tabular}{|c|c|c|c|c|c|c|c|c|}\hline
$N$ & 1 & 2 & 3 & 4 & 5 & 6 & 7 & 8 \\ \hline
$\rho$ & $1^{24}$ & $1^82^8$ & $1^63^6$ & $1^42^24^4$ & $1^45^4$ & $1^22^23^26^2$ & $1^37^3$ & $1^22^14^18^2$ \\ \hline
cycle sum & 24 & 16 & 12 & 10 & 8 & 8 & 6 & 6 \\ \hline
\end{tabular}
\caption{ \sf Cycle shapes for $\mathbb{Z}_N$ orbifolds of $ \text{K3} \times \mathrm{S}^1 \times \widetilde{\mathrm{S}}^1$. }
\label{table_2}
\end{center}
\end{table}

\begin{table}[t]
\begin{center}
\begin{tabular}{|c|c|c|c|}\hline
$N$  & 2 & 3 & 4  \\ \hline
$\rho$  & $1^{-8}2^{16}$ & $1^{-3}3^{9}$ & $1^{-4}2^64^{4}$  \\ \hline
\end{tabular}
\caption{\sf Frame shapes for  $\mathbb{Z}_N$ orbifolds of $\text{T}^4 \times \mathrm{S}^1 \times \widetilde{\mathrm{S}}^1$ in the D1-D5 frame.}
\label{table3}
\end{center}
\end{table} 

\subsection{Hair partition functions}
\label{hair_partition_functions}

\paragraph{5d hair partition functions:} 
A set of hair of the 5d black hole consists of 12
gravitino zero modes corresponding to 12 broken supersymmetries. 
Since the four unbroken supersymmetries are left-chiral, eight of the broken supersymmetries are right-chiral and four of the broken supersymmetries are left-chiral.  These zero modes give a contribution to
the partition function of the form
$(e^{\pi i \widetilde v}-e^{-\pi i \widetilde v})^4$~\cite{Banerjee:2009uk}.

We found in section \ref{sec:fermionic_deformations} that there are 4 left-moving  gravitino hair modes. These modes are the uplift of the four left-chiral zero modes mentioned in the previous paragraph. These modes give additional contribution to the partition function of the black hole hair.  

Since these modes are periodic in our six-dimensional supergravity description, these modes can only carry integer units of momentum along the orbifolded circle $\mathrm{S}^1/\mathbb{Z}_N$. In particular, these modes do not  carry the most general momentum allowed quantised in units of $1/N$. (Say, a mode carries $l \in \mathbf{Z}_{+}$  integer units of left momentum along the orbifolded circle $\mathrm{S}^1/\mathbb{Z}_N$, then since $-l = \frac{1}{N} (-N l)$, it carries $N l$ units of momentum in our conventions.) As a result, 
these modes give a contribution to the hair partition function,
\be
\prod_{l=1}^\infty \, (1-e^{2\pi i N l \widetilde \rho})^4.
\ee
Combining these
two contributions we get
\be 
Z^\rom{hair}_\rom{5d}(\widetilde \rho,\widetilde \sigma, \widetilde v)
=  (e^{\pi i \widetilde v}-e^{-\pi i \widetilde v})^4\,
\prod_{l=1}^\infty \, (1-e^{2\pi i N l \widetilde \rho})^4 \, .
\ee
For $N=1$ this answer is the same as the one given in \cite{Jatkar:2009yd}. The hair removed partition function $Z_\rom{5d}^\rom{hor}(\widetilde \rho,\widetilde \sigma, \widetilde v)$ is 
\be
Z_\rom{5d}^\rom{hor}(\widetilde \rho,\widetilde \sigma, \widetilde v) = \frac{Z_\rom{5d}(\widetilde \rho,\widetilde \sigma, \widetilde v)}{Z^\rom{hair}_\rom{5d}(\widetilde \rho,\widetilde \sigma, \widetilde v)} = \frac{f(\widetilde \rho)}{\widetilde\Phi(\widetilde\rho,\widetilde\sigma,\widetilde v)}  e^{-2 \pi i \widetilde \rho}  \prod_{l=1}^\infty \, (1-e^{2\pi i N l \widetilde \rho})^{-4}. 
\ee

\paragraph{4d hair partition functions:} 

In this case, the hair modes include 12 fermion zero modes.  By construction, they are all used in saturating the helicity factors inserted
into the helicity trace. Hence these zero modes simply contribute 1~\cite{Banerjee:2009uk, Jatkar:2009yd}.  Besides these, there are $n_t$ left-moving
bosonic modes associated with the 2-form deformations and
3 left-moving bosonic modes associated with the transverse oscillations
of the black hole. All these modes are neutral under the $x^4$
translation. Finally, as in the 5d case, there are four left-moving gravitino modes,
also neutral under $x^4$. These four fermionic modes cancel the
contribution from four of the bosonic modes. Since these modes are periodic, they only carry integer quantised momentum, $l \in \mathbf{Z}_{+}$. Thus, their contribution to the hair partition function is:
\be  
Z_\rom{4d}^\rom{hair}(\widetilde\rho,\widetilde\sigma,\widetilde v) = \prod_{l=1}^\infty
\left(1 - e^{2\pi i N l \widetilde\rho}\right)^{-n_t +1} \,.
\ee
For $n_t = 21$ and $N=1$ this answer is the same as the one given in \cite{Jatkar:2009yd}. For $N \neq 1$ this is not the end of the story. There are additional hair modes. They come from the twisted sectors. A way to incorporate the twisted sectors in supergravity  is to analyse the problem in ten-dimensions.

Let us schematically denote   $y$ to be the $\text{K3}$ directions and $x$ to be the remaining six-dimensions  in the \emph{unorbifolded} theory. Then, in ten-dimensions the RR four-form field schematically decomposes as~\cite{hep-th/9506126},
\be
\hat C_4(x,y) = C_4(x) + \sum_{\gamma} c^\gamma_{2}(x) \wedge \omega^\gamma(y) + c_{0} (x) \star_\text{K3} \mathbb{1} (y),
\ee
where  $\omega^\gamma(y)$ are the self-dual and anti-self-dual harmonic forms spanning the cohomology $H^2(\text{K3}).$ On the elements on this cohomology,  the abelian orbifold group of order $N$  generated by $\widetilde{g}$ acts.

To obtain the six-dimensional supergravity description discussed in the previous sections only the $\widetilde{g}-$invariant forms were kept. In ten-dimensions, however,  a more general situation is possible, where $\omega^\gamma(y)$ are not $\widetilde{g}-$invariant and accordingly the fields $c^\gamma_{2}(x)$ must pick up the \emph{opposite}  phase under the orbifold action. This phase comes because the orbifold action also  involves a shift along the S$^1$.  The combined effect ensures that the ten-dimensional $\hat C_4(x,y) $ is $\widetilde{g}-$invariant. These modes give rise to additional hair modes.\footnote{Since $\hat C_4(x,y) $ in ten-dimension is $\widetilde{g}-$invariant, nothing special is needed to analyse the smoothness of these modes in ten-dimensions. We expect that the discussion of section \ref{sec:bosonic_deformations_TN_form_fields} admits a straightforward extension to ten-dimensions. We leave the precise details to be worked out in the future. \label{footnote:twisted_hair}} 
In order to account for their contributions to the partition functions, we want to know the number of harmonic 2-forms $\omega^\gamma(y)$ transforming as 
$\widetilde{g}^s$ for $0 \le s \le N-1$. From this information we would know the hair modes that are \emph{not} neutral under the orbifold action,  but must satisfy the boundary condition 
\be
c^\gamma_{2}(t, x^5 + 2\pi, x^4, r, \theta, \phi) = \exp\left[- 2 \pi i \frac{s}{N} \right]c^\gamma_{2}(t, x^5, x^4, r, \theta, \phi). \label{twisted_BC}
\ee

For the CHL models this information can be read from~\cite{Chaudhuri:1995dj}. This data is summarised in table \ref{table4}. 

Let us start our discussion for accounting for these modes for $N=2$. For $N=2$ there are  8 additional hair modes with anti-periodic boundary conditions (for $N=2$ only non-trivial choice in $s=1$ in equation \eqref{twisted_BC}): 
\be
c^\gamma_{2}(t, x^5 + 2\pi, x^4, r, \theta, \phi) = - c^\gamma_{2}(t, x^5, x^4, r, \theta, \phi).
\ee

 Since these modes are anti-periodic along the orbifolded circle $\mathrm{S}^1/\mathbb{Z}_2$,  they carry odd units of momentum $-n = - (2k - 1)$ along in the unorbifolded circle $\mathrm{S}^1$. Thus, their contribution to the partition function to the $\mathbb{Z}_2$ CHL model is, 
\be
 \prod_{k=1}^\infty
\left(1 - e^{2\pi i  (2k - 1) \widetilde\rho}\right)^{-8} = \left(1 - q\right)^{-8} \cdot \left(1 - q^3\right)^{-8} \cdot \left(1 - q^5\right)^{-8} \ldots.
\ee
where \be
q = e^{2\pi i \widetilde\rho}.
\ee Together with the contribution from the periodic modes we get
\bea
Z_\rom{4d}^\rom{hair}(\widetilde\rho,\widetilde\sigma,\widetilde v) &=& 
\left(1 - q^2\right)^{-12} \cdot \left(1 - q^4\right)^{-12} \cdot \left(1 - q^6\right)^{-12} \ldots 
\\ && ~\times \left(1 - q\right)^{-8} \cdot \left(1 - q^3\right)^{-8} \cdot \left(1 - q^5\right)^{-8} \ldots
\eea
The hair removed partition function $Z_\rom{4d}^\rom{hor}(\widetilde \rho,\widetilde \sigma, \widetilde v)$ is 
\bea
Z_\rom{4d}^\rom{hor}(\widetilde \rho,\widetilde \sigma, \widetilde v) &=& \frac{Z_\rom{4d}(\widetilde \rho,\widetilde \sigma, \widetilde v)}{Z^\rom{hair}_\rom{4d}(\widetilde \rho,\widetilde \sigma, \widetilde v)} 
 = \frac{1}{\widetilde\Phi(\widetilde\rho,\widetilde\sigma,\widetilde v)}  \times
\left(1 - q^2\right)^{12} \cdot \left(1 - q^4\right)^{12} \cdot \left(1 - q^6\right)^{12} \ldots 
\nn \\ && ~\times \left(1 - q\right)^{8} \cdot \left(1 - q^3\right)^{8} \cdot \left(1 - q^5\right)^{8} \ldots
\eea

 \begin{table}[t!]
\begin{center}
\begin{tabular}{|c||c|c|c|c|c|c|c|c|c|} \hline
$N$ & $b_0$ & $b_1$ & $b_2$ & $b_3$ & $b_4$ & $b_5$ & $b_6$ & $b_7$  & $n_t + 3 = b_0 + 5$\\ \hline \hline
1&19&  &  & & &  & & & 24  \\ \hline
2&11&8&  & & &  & &  & 16 \\ \hline
3&7&6&6&  &  &  & &  & 12 \\ \hline
4&5 &4&6  &4  &  & &  &  &10 \\ \hline
5&3  &4 &4 &4 &4 &  & & &   8\\ \hline
6&3 &2 &4  &4  &4  &2 &  &  &8 \\ \hline
7&1 &3 &3 &3 &3 &3 &3 & &  6\\ \hline
8&1 &2 &3  &2  &4   &2 &3  &2  & 6 \\ \hline
\end{tabular}
\caption{ \sf  Hodge data on $\mathbb{Z}_N$ orbifolds of $ \text{K3} \times \mathrm{S}^1 \times \widetilde{\mathrm{S}}^1$. The numbers $b_s$ denote the number of anti-self-dual $(1,1)$ form transforming as $\widetilde{g}^s$. We note that number of tensor multiplets in  six-dimensional supergravity description is simply the number of $\widetilde{g}$ invariant anti-self-dual (1,1) forms $b_0$ plus 2: $n_t = b_0 + 2$. The plus 2 comes from the self-dual and anti-self-dual decomposition of the type IIB RR and NS-NS 2-form fields. The ratio of the 4d and 5d partition functions depends on $n_t+3$, which is equal to $b_0 + 5$. This information is listed in the last column. }
\label{table4}
\end{center}
\end{table}

\subsection{Matching of the 4d/5d horizon partition functions}
\label{matching}
From the previous discussion we have, for $N=2$ $f(\widetilde \rho) = \eta (2 \widetilde \rho)^8\eta ( \widetilde \rho)^8$. Thus, 
\bea
\frac{Z_\rom{5d}^\rom{hor}}{Z_\rom{4d}^\rom{hor}} &=&  f(\widetilde \rho) \, e^{-2 \pi i \widetilde \rho} 
\left(1 - q^2\right)^{-16} \cdot \left(1 - q^4\right)^{-16} \cdot \left(1 - q^6\right)^{-16} \ldots 
\\ && ~\times \left(1 - q\right)^{-8} \cdot \left(1 - q^3\right)^{-8} \cdot \left(1 - q^5\right)^{-8} \ldots \\
&=&  f(\widetilde \rho) \, e^{-2 \pi i \widetilde \rho} \prod_{k=1}^{\infty} \left(1 - q^{2k}\right)^{-8} \cdot \prod_{l=1}^{\infty} \left(1 - q^l\right)^{-8} \\
&=&1.
\eea
The horizon partition functions  perfectly match.\footnote{In reference \cite{Banerjee:2009uk} the small black hole contributions to the partition functions were also considered. Following \cite{Jatkar:2009yd}, we ignore this complication in this paper.}

This is not a coincidence. Using information from table \ref{table4} and appropriate periodicity of the modes we see that the hair removed partition functions match in all cases.  For $N=1$, $n_t = 21$, $f(\widetilde \rho) = \eta^{24}(\widetilde \rho)$. This matching was already observed in \cite{Jatkar:2009yd}. Let us then check for the rest of the values of $N$, i.e., $N=3,4,5,6,7,8$ one by one.

The basic ingredients for this check is to note that for a given value of $N$, the contribution to the 4d hair partition function due to the modes transforming as $\widetilde{g}^s$ is of the form,
\begin{equation}
    Z^{\text{hair}}_{\text{4d}, b_s} = \prod_{l=1}^{\infty} (1 - q^{lN-s})^{-(b_s + 5 \delta_{s,0})} \;\;\; ,\;\;s =0,1,\cdots N-1,
\end{equation}
with the full $4$d hair partition function given by the product,
\begin{equation}
    Z^{\text{hair}}_{\text{4d}} = \prod_{s=0}^{N-1} Z^{\text{hair}}_{\text{4d}, b_s} \;\;\;.
\end{equation}
Also it is useful to keep in mind the identity,
\begin{equation}
    \eta(N \Tilde{\rho}) = q^{N/24} \prod_{l=1}^{\infty} (1- q^{Nl}) \;\;.
\end{equation}
\paragraph{\textbf{\underline{Case of $N=3$:}}} For this case $ f(\widetilde\rho) = \eta(\widetilde\rho)^{6} \eta(3\widetilde\rho)^{6}$.
From table \ref{table4} we see there are six  modes that satisfy
\be
c^\gamma_{2}(t, x^5 + 2\pi, x^4, r, \theta, \phi) = \exp\left[- \frac{2 \pi i}{3} \right]c^\gamma_{2}(t, x^5, x^4, r, \theta, \phi),
\ee
and six modes that satisfy
\be
c^\gamma_{2}(t, x^5 + 2\pi, x^4, r, \theta, \phi) = \exp\left[- \frac{4 \pi i}{3}\right]c^\gamma_{2}(t, x^5, x^4, r, \theta, \phi). 
\ee 
The ratio of the 5d and 4d partition functions is therefore, 
\bea
\frac{Z_\rom{5d}^\rom{hor}}{Z_\rom{4d}^\rom{hor}} &=&  f(\widetilde \rho) \, e^{-2 \pi i \widetilde \rho} 
\left(1 - q^3\right)^{-12} \cdot \left(1 - q^6\right)^{-12} \cdot \left(1 - q^9\right)^{-12} \ldots \nn \\
&& \times 
\left(1 - q\right)^{-6} \cdot \left(1 - q^4\right)^{-6} \cdot \left(1 - q^7\right)^{-6} \ldots \nn \\
&& \times 
\left(1 - q^2\right)^{-6} \cdot \left(1 - q^5\right)^{-6} \cdot \left(1 - q^8\right)^{-6} \ldots \\
&=&  f(\widetilde \rho) \, e^{-2 \pi i \widetilde \rho}  \prod_{k=1}^{\infty} \left(1 - q^{3k}\right)^{-6} \cdot \prod_{l=1}^{\infty} \left(1 - q^l\right)^{-6} \\
&=&1.
\eea
\paragraph{\textbf{\underline{Case of $N=4$:}}} For this case $ f(\widetilde\rho) = \eta(\widetilde\rho)^{4} \eta(2\widetilde\rho)^{2} \eta(4\widetilde\rho)^{4}$. Once again using the entries from table \ref{table4} we can express the ratio of the $5$d and $4$d partition function as,
\bea
\frac{Z_\rom{5d}^\rom{hor}}{Z_\rom{4d}^\rom{hor}} &=&  f(\widetilde \rho) \, e^{-2 \pi i \widetilde \rho} 
\left(1 - q^4\right)^{-10} \cdot \left(1 - q^8\right)^{-10} \cdot \left(1 - q^{12}\right)^{-10} \ldots \nn \\
&& \times 
\left(1 - q\right)^{-4} \cdot \left(1 - q^5\right)^{-4} \cdot \left(1 - q^{9}\right)^{-4} \ldots \nn \\
&& \times 
\left(1 - q^2\right)^{-6} \cdot \left(1 - q^6\right)^{-6} \cdot \left(1 - q^{10}\right)^{-6} \ldots \nn \\
&& \times 
\left(1 - q^3\right)^{-4} \cdot \left(1 - q^7\right)^{-4} \cdot \left(1 - q^{11}\right)^{-4} \ldots \\
&=&  f(\widetilde \rho) \, e^{-2 \pi i \widetilde \rho}  \prod_{k=1}^{\infty} \left(1 - q^{4k}\right)^{-4} \cdot \prod_{l=1}^{\infty} \left(1 - q^l\right)^{-4} \cdot \prod_{j=1}^{\infty} \left(1 - q^{2j}\right)^{-2}  \\
&=&1.
\eea

\paragraph{\textbf{\underline{Case of $N=5$:}}}  For this case $ f(\widetilde\rho) = \eta(\widetilde\rho)^{4} \eta(5\widetilde\rho)^{4}$.
Reading entries of table \ref{table4}, the ratio can be expressed as

\bea
\frac{Z_\rom{5d}^\rom{hor}}{Z_\rom{4d}^\rom{hor}} &=&  f\left(\widetilde \rho\right) \, e^{-2 \pi i \widetilde \rho} \nonumber\\
 &\times& \prod_{k=1}^{\infty} \bigg[ \left(1-q^{5k}\right)^{-4} \left(1-q^{5k}\right)^{-4} \left(1-q^{5k-1}\right)^{-4} \nn\\
 && \times \left(1-q^{5k-2}\right)^{-4} \left(1-q^{5k-3}\right)^{-4} \left(1-q^{5k-4}\right)^{-4}\bigg] \nonumber \\
&=& f\left(\widetilde \rho\right) \, e^{-2 \pi i \widetilde \rho} \prod_{k=1}^{\infty} \bigg[\left(1-q^{5k}\right)^{-4} \left(1-q^{k}\right)^{-4}\bigg] \nn\\
&=& 1.
\eea

\paragraph{\textbf{\underline{Case of $N=6$:}}} For $N=6$, $ f(\widetilde\rho) = \eta(\widetilde\rho)^{2} \eta(2\widetilde\rho)^{2} \eta(3\widetilde\rho)^{2} \eta(6\widetilde\rho)^{2}. $
Repeating the same procedure as before,
\bea
\frac{Z_\rom{5d}^\rom{hor}}{Z_\rom{4d}^\rom{hor}} &=&  f\left(\widetilde \rho \right) \, e^{-2 \pi i \widetilde \rho} \nonumber\\
 &\times& \prod_{k=1}^{\infty} \bigg[ \left(1-q^{6k}\right)^{-4} \left(1-q^{6k}\right)^{-4} \left(1-q^{6k-1}\right)^{-2} \left(1-q^{6k-2}\right)^{-4} \nn \\
 &&\times\left(1-q^{6k-3}\right)^{-4} \left(1-q^{6k-4}\right)^{-4} \left(1-q^{6k-5}\right)^{-2}\bigg] \nonumber \\
&=& f\left(\widetilde \rho\right) \, e^{-2 \pi i \widetilde \rho} \prod_{k=1}^{\infty} \bigg[\left(1-q^{k}\right)^{-2} \left(1-q^{2k}\right)^{-2} \left(1-q^{3k}\right)^{-2}  \left(1-q^{6k}\right)^{-2}  \bigg] \nn\\
&=& 1.
\eea

\paragraph{\textbf{\underline{Case of $N=7$:}}}  For this case $ f(\widetilde\rho) = \eta(\widetilde\rho)^{3} \eta(7\widetilde\rho)^{3}$.
This case addresses the last of the non-composite values of $N$. From table \ref{table4}, we can write
\bea
\frac{Z_\rom{5d}^\rom{hor}}{Z_\rom{4d}^\rom{hor}} &=&  f\left(\widetilde \rho\right) \, e^{-2 \pi i \widetilde \rho} \nonumber\\
 &\times& \prod_{k=1}^{\infty} \bigg[ \left(1-q^{7k}\right)^{-3} \left(1-q^{7k}\right)^{-3} \left(1-q^{7k-1}\right)^{-3} \left(1-q^{7k-2}\right)^{-3} \nn \\
 &&\times\left(1-q^{7k-3}\right)^{-3} \left(1-q^{7k-4}\right)^{-3} \left(1-q^{7k-5}\right)^{-3} \left(1-q^{7k-6}\right)^{-3}\bigg] \nonumber \\
&=& f\left(\widetilde \rho\right) \, e^{-2 \pi i \widetilde \rho} \prod_{k=1}^{\infty} \bigg[\left(1-q^{7k}\right)^{-3} \left(1-q^{k}\right)^{-3} \bigg] \nn\\
&=& 1.
\eea

\paragraph{\textbf{\underline{Case of $N=8$:}}}
For $N =8$, $f(\widetilde\rho) = \eta(\widetilde\rho)^{2} \eta(2\widetilde\rho) \eta(4\widetilde\rho)  \eta(8\widetilde\rho)^{2}$,  and we have
\bea
\frac{Z_\rom{5d}^\rom{hor}}{Z_\rom{4d}^\rom{hor}} &=&  f\left(\widetilde \rho\right) \, e^{-2 \pi i \widetilde \rho} \nonumber\\
 &\times& \prod_{k=1}^{\infty} \bigg[ \left(1-q^{8k}\right)^{-4} \left(1-q^{8k}\right)^{-2} \left(1-q^{8k-1}\right)^{-2} \left(1-q^{8k-2}\right)^{-3} \nn \\
 &&\times\left(1-q^{8k-3}\right)^{-2} \left(1-q^{8k-4}\right)^{-4} \left(1-q^{8k-5}\right)^{-2} \left(1-q^{8k-6}\right)^{-3} \left(1-q^{8k-7}\right)^{-2}\bigg] \nonumber \\
&=& f\left(\widetilde \rho\right) \, e^{-2 \pi i \widetilde \rho} \prod_{k=1}^{\infty} \bigg[\left(1-q^{k}\right)^{-2} \left(1-q^{2k}\right)^{-1} \left(1-q^{4k}\right)^{-1}  \left(1-q^{8k}\right)^{-2}  \bigg] \nn\\
&=& 1.
\eea

Thereby we have explicitly shown that the 4d and 5d horizon partition functions match perfectly for  $\text{K3}/\mathbb{Z}_N$ CHL models for $N=1,2,3,4,5,6,7,8$.

\section{Discussion}
\label{sec:conclusions}

In this paper, we have  presented  hair modes in the untwisted as well as twisted sectors for a wide class of  CHL models.  We have shown that  after removing the contributions of the hair modes from the microscopic partition functions,  the 4d and 5d horizon partition functions match. We have also presented  details on the smoothness analysis of hair modes for rotating black holes, which were largely missing from the literature.

Our results offer several opportunities for future research. Perhaps the most interesting among them is an analysis of hair modes for the T$^4$ orbifold models with ${\cal N} = 4$ supersymmetry.  It will be most convenient to analyse this problem in ten-dimensional type IIB supergravity. For these set-ups the dynamics of Wilson lines along T$^4$ also contributes to the microscopic partition functions \cite{hep-th/0609109, 0708.1270}. At the same time, there is a possibility of additional hair modes (with excitations along the T$^4$ directions) \cite{Mishra:2018bcb, Chakrabarti:2019lfu}. It will be interesting to analyse this problem and understand the hair removed partition functions.   These models were also  recently studied in reference \cite{Chattopadhyaya:2018xvg}, where the authors noted that  the sign of the index for sufficiently low charges for single-center 4d black holes violates the positivity conjecture of \cite{Sen:2010mz}.  The hair removed partition functions are expected to satisfy the positivity conjecture. It will be  interesting to check this explicitly.

Although not analysed in this paper, it is expected from the analysis of \cite{hep-th/0609109, 0708.1270} that the matching we showed above also works for more general  $\mathbb{Z}_N$ CHL orbifold models (non-geometric orbifolds).  As we saw in section \ref{sec:microscopics}, the agreement between 
the 4d and 5d hair removed partition functions 
essentially boils down to consistency between table  \ref{table1} and \ref{table4}. It was shown in \cite{hep-th/0609109, 0708.1270, Govindarajan:2011mp} that the Hodge data for   orbifolds for K3 (and also for T4) is directly related to the data that enters  in the construction of the $\frac{1}{2}$-BPS partition functions. It will be interesting to  revisit this in the context of hair removal.

The smoothness analysis of twisted sector hair modes requires a ten-dimensional discussion. As mentioned in footnote \ref{footnote:twisted_hair}, we expect the details to be a straightforward extension of the analysis in section
\ref{sec:bosonic_deformations_TN_form_fields}. It will be useful to work out these details in the future.

The puzzle of the difference between the 4d and 5d partition functions seems to be even more challenging for ${\cal N}=8$ compactification~\cite{Maldacena:1999bp, Shih:2005qf, Sen:2008ta} than the cases analysed in this paper. We wonder if the hair modes can account for the difference. 

We hope to report on some of these problems in our future work.

\subsection*{Acknowledgements} We thank Sujay Ashok, Justin David, Anshuman Maharana, Samir Mathur, Swapnamay Mondal,  and especially  Nabamita Banerjee, Dileep Jatkar,  Bindusar Sahoo, and Ashoke Sen for helpful discussions on various aspects of this work. AV and PS additionally thank Bindusar Sahoo for warm hospitality at IISER TVM during the initial stages of this work and for numerous discussions on supergravity theories. AV also thanks YKS for warm hospitality at NISER Bhubaneswar. The work of AV and PS is supported in part by the Max Planck Partnergroup ``Quantum Black Holes'' between CMI Chennai and AEI Potsdam and by a grant to CMI from the Infosys Foundation.

\appendix

\section{The notion of weight}
\label{app:weights}
The authors of \cite{Jatkar:2009yd} introduced the concept of weight that proves very convenient, both in the analysis of the background solutions and  in the analysis of the bosonic and fermionic deformations. Here we review this concept and expand on it.

The notion of weight is defined for \textit{components} of any tensor in $u$ and $v$ (or  $U$ and $V$) coordinates. The tensor can be contravariant, covariant, or  mixed.  For a given component of a covariant tensor, weight is defined as 
\begin{equation}
    wt_{_{cov}} = \# \text{ of } v \text{ indices } - \# \text{ of } u \text{ indices}.
\end{equation}
For a given component of a contravariant tensor, weight is defined as 
\begin{equation}
    wt_{_{cont}} = \# \text{ of } u \text{ indices } - \# \text{ of } v \text{ indices},
\end{equation}
and for a given component of a generic mixed tensor, weight is defined as
\begin{equation}
    wt = wt_{_{cov}} + wt_{_{cont}} \,.
\end{equation}
Note that the notion of weight is exactly opposite for the contravariant and covariant tensor components. 

Given any tensor, we can think of it as a collection of its components. Each  component, by definition, has a unique weight assigned to it. Therefore given a tensor, we can  always write it as a tensor sum of tensor components of fixed weight. For example, consider a rank-2 covariant tensor $P_{MN}$. Let us decompose any index as $M = \lbrace u\,,v\,,i\, \rbrace$. 
    Then we can write,
    \begin{equation}
         P_{M N} = \underbrace{P_{u u}}_{\textit{weight }-2} \bigoplus \underbrace{P_{u i} \oplus P_{i u}}_{\textit{weight }-1} \; \bigoplus \; \underbrace{P_{uv} \oplus P_{v u} \oplus  P_{i j}}_{\textit{weight }0} \;\bigoplus\; \underbrace{P_{v i} \oplus P_{i v}}_{\textit{weight }1} \bigoplus \underbrace{P_{v v}}_{\textit{weight }2}
    \end{equation}
Generically, a rank-$r$ covariant tensor decomposes as
\begin{equation}
        P_{{M_i}\dots {M_r}} = \bigoplus_{w=-r}^{w=r} \Bigg[\bigoplus_{wt(\lbrace M_i\rbrace)=w} P_{\lbrace M_i \rbrace}\Bigg].
\end{equation}
Since the definition of the weight is exactly opposite for a covariant and contravariant index, under index contraction the weight remains unchanged.  As an example, consider  $P_{M N} = Q_{M N T}^{\qquad \; \; T}$. We can write, 
     \be
     P_{M N} = Q_{M N v}^{\qquad \; \; v} + Q_{M N u}^{\qquad \; \; u} + Q_{M N i}^{\qquad \; \; i}.
     \ee 
     Each term on RHS has the same weight as the LHS for a given $M, N$.

The definition of weight, while dependent on the specific choice of coordinates $u$ and $v$, is completely insensitive to the choice of coordinates on the transverse 4d space.  The weight of a given component of a tensor, remains the same whether we use $w^i$ coordinates or any other coordinates, such as $(r,\, \theta,\, \phi,\, x^4)$ to describe the 4d transverse space.

Throughout our paper, we follow the convention that derivatives are with respect to contravariant coordinates only, i.e., derivatives are always with lower indices. A rule of thumb is then: $\partial_ u$ decreases weight by 1 and $\partial_ v$ increases weight by 1. Note that taking derivatives with respect to $u$ or $v$, or multiplying with other  tensors, or multiplying with objects like gamma matrices or spin connection coefficients, are the only ways to change the weight of a given tensor component.

 \section{Action of Lorentz boosts and null rotations on gravitino configurations}
\label{app:Lorentz}
In the main text, at two places we came across the following Lorentz transformations in six-dimensions,
\begin{align}
& \hat e^+ = \alpha e^+, &
& \hat e^- = \frac{1}{\alpha} e^-, 
& \hat e^2 = e^2, &
& \hat e^3 = e^3,
\end{align}
followed by, 
\begin{align}
& \check e^+ = \hat e^+,&
& \check e^- = \hat e^- - 2 \beta \hat  e^3 - \beta^2 \hat e^+,
& \check e^2 = \hat e^2, &
& \check e^3 = \hat  e^3 + \beta \hat e^+,
\end{align}
and (optionally), 
\begin{align}
& \tilde e^+= \check e^+ ,&
& \tilde e^- = \check e^- - 2 \gamma \check e^2 - \gamma^2 \check e^+,
& \tilde e^2 = \check e^2 + \gamma \check e^+ .&
& \tilde e^3 = \check e^3.
\end{align}

In this appendix we discuss the action of the above Lorentz boost (parametrised by $\alpha)$ and null rotations (parametrised by $\beta$ and $\gamma)$ on gravitinos configurations of interest. These results were used in \cite{Jatkar:2009yd}. The various minus signs can be a potential source of confusion, so we work this out explicitly.  Let us start with the boost. The boost can be written as 
\be
\hat e^A = \Lambda^A{}_B \, e^B, \quad \text{with} \quad \Lambda^A{}_B = \left( \exp \left [ \frac{1}{2} \lambda^{CD} m_{[CD]}\right] \right) ^A{}_B,
\ee
with Lorentz generators in six-dimensional vector representation,
\be
m_{[CD]} {}^A{}_B = \delta^A_C \eta_{BD} - \delta^A_D \eta_{CB} = - m_{[DC]} {}^A{}_B,
\ee
and parameters $\lambda^{01} = - \lambda^{10} = \log \alpha$.

We choose the following gamma matrices in six-dimensions (as in the main text), 
\begin{align}
\widetilde \Gamma^0 &=   \, \, \,  \mathbb{1}_2 \otimes   \mathbb{1}_2 \otimes (-\mathrm{i}) \sigma_1, &
\widetilde \Gamma^1 &= \, \, \, \mathbb{1}_2 \otimes  \mathbb{1}_2 \otimes \sigma_2, \\
\widetilde \Gamma^2 &=  \, \, \, \mathbb{1}_2\otimes  \sigma_1 \otimes  \sigma_3, &
\widetilde \Gamma^3 &=  \, \, \, \sigma_3 \otimes  \sigma_3 \otimes  \sigma_3, \\
\widetilde \Gamma^4 &=  \, \, \, \sigma_1 \otimes  \sigma_3 \otimes  \sigma_3, &
\widetilde \Gamma^5 &=  \, \, \, \sigma_2 \otimes  \sigma_3 \otimes  \sigma_3.
\end{align}

The gravitino configurations of interest are of the form $\Psi_M = 0 $ for $M \neq V$, together with the projection condition $(\widetilde \Gamma^0 + \widetilde \Gamma^1)  \Psi_V =\widetilde \Gamma^+ \Psi_V = 0$. These conditions imply that the field $\Psi_V$ is of the general form, 
\be
\Psi_V = \{ \psi_1, 0, \psi_3, 0, \psi_5, 0, \psi_7,0\}^T, \label{spinor_general_form}
\ee
where $\psi_1, \psi_3, \psi_5, \psi_7$ denote the general non-zero entries. 

The action of Lorentz boosts on the spinor $\Psi_V$ is simply,
\be
\exp \left[ \frac{1}{2} \lambda^{01}  \Sigma_{01}  +  \frac{1}{2} \lambda^{10}  \Sigma_{10}   \right] \Psi_V
\ee
where
\be
\Sigma^{AB} = \frac{1}{4} [  \widetilde \Gamma^A, \widetilde \Gamma^B ]. 
\ee
A short calculation shows that 
\be
\exp \left[ \frac{1}{2} \lambda^{01}  \Sigma_{01}  +  \frac{1}{2} \lambda^{10}  \Sigma_{10}   \right] \Psi_V = e^{-\frac{1}{2}\log \alpha}  \Psi_V = \frac{1}{\sqrt{\a}}  \Psi_V.
\ee
Thus, the Lorentz boost simply acts as a rescaling. For our applications $ \alpha \propto -2  V$, which gives the result that in the hatted Lorentz frame,
\be
\widehat{\Psi}_V \propto (-2V)^{-1/2}  \Psi_V.
\ee

The null rotation parametrised by $\beta$, 
$
\check e^A = \Lambda^A{}_B \, \hat e^B, 
$
is generated with $\lambda^{03} = \beta$ and $\lambda^{13} = - \beta$.  
Since $ \left( \beta  \Sigma_{03}  - \beta  \Sigma_{13} \right)  \propto \beta \widetilde{\Gamma}^{3+}  $, it follows that for a spinor annihilated by $\widetilde{\Gamma}^{+}$,  
\be
 \exp \left[ \beta  \Sigma_{03}  - \beta  \Sigma_{13}   \right] \widehat{\Psi}_V = \mathbb{1} \cdot  \widehat{\Psi}_V = \widehat{\Psi}_V.
\ee
That is, such a null rotation does not change the spinor. 

The null rotation parametrised by $\gamma$ is generated with $\lambda^{02} = \gamma$ and $\lambda^{12} = - \gamma$. Once again  such a null rotation does not change the spinor annihilated by $\widetilde{\Gamma}^{+}$.

\end{document}